\documentclass[onecolumn,11pt,aps,prd,superscriptaddress,nofootinbib,longbibliography]{revtex4-1}
\usepackage{amsmath,amssymb}
\usepackage[colorlinks,linkcolor=red,anchorcolor=blue,citecolor=cyan]{hyperref} 
\usepackage{graphicx}
\usepackage{adjustbox}
\usepackage{tikz}
\usetikzlibrary{shadings}
\usetikzlibrary{arrows.meta}
\usetikzlibrary{calc}
\usepackage{colortbl}
\usepackage{xcolor}
\usepackage{picinpar}
\usepackage{float}
\usepackage{wrapfig}

\usepackage{slashed}
\usepackage{ulem}
\usepackage{bm}
\usepackage{mathrsfs}
\usepackage{array,cases}
\usepackage{comment}
\usepackage{multirow,paracol}

\usepackage[final]{changes}
\usepackage{lipsum}

\usepackage{amsmath,mathtools,amssymb}
\usepackage{bm,extarrows,ulem,cancel,slashed}
\usepackage{mathrsfs}
\usepackage{esint}
\usepackage{geometry,graphicx,color}
\usepackage[all,pdf]{xy}
\usepackage{pstricks,pstricks-add}
\usepackage{hyperref}
\hypersetup{pdfborder=001,colorlinks=true, linkcolor=black}
\usepackage{listings}

\newcommand{\be}{\begin{equation}}
\newcommand{\ee}{\end{equation}}
\newcommand{\bea}{\begin{eqnarray}}
\newcommand{\eea}{\end{eqnarray}}
\newcommand{\ba}{\begin{array}}
\newcommand{\ea}{\end{array}}
\newcommand{\bs}{\be\begin{split}}
\newcommand{\es}{\end{split}}

\newcommand{\dif}{\,\mathrm{d}}
\newcommand{\mi}{\mathrm{i}}
\newcommand{\me}{\mathrm{e}}

\newcommand{\p}{\partial}
\renewcommand{\1}{\left}
\renewcommand{\2}{\right}
\newcommand{\ma}{\mathcal}
\newcommand{\ri}{\rightarrow}

\newcommand{\rom}[1]{\uppercase\expandafter{\romannumeral #1}}

\newcommand{\m}{\mu}
\newcommand{\n}{\nu}
\newcommand{\al}{\alpha}

\newcommand{\ep}{\epsilon}

\newcommand{\del}{\delta}

\renewcommand{\th}{\theta}

\newcommand{\bes}{\be\1\{\begin{split}}

\def\roughly#1{\mathrel{\raise.3ex\hbox{$#1$\kern-.75em%
\lower1ex\hbox{$\sim$}}}}

\def\({\left(}
\def\){\right)}
\def\[{\left[}
\def\]{\right]}
\def\<{\langle}
\def\>{\rangle}

\def\k{{\kappa}}
\def\l{{\lambda}}

\def\d{{\delta}}
\def\D{{\Delta}}
\def\o{{\omega}}
\def\O{{\Omega}}

\def\a{{\alpha}}
\def\b{{\beta}}

\def\G{{\Gamma}}

\def\m{{\mu}}
\def\n{{\nu}}
\def\r{{\rho}}

\def\t{{\tau}}
\def\th{{\theta}}

\def\x{{\xi}}

\allowdisplaybreaks

\begin{document}

\author{Nayun Jia}
\email{nayun.jia@foxmail.com}
\affiliation{Department of Physics, Southern University of Science and Technology, Shenzhen 518055, China}
\affiliation{Key Laboratory of Cosmology and Astrophysics (Liaoning) \& College of Sciences, Northeastern University, Shenyang 110819, China}

\author{Yin-Da Guo}
\email{yinda.guo@mail.sdu.edu.cn}
\affiliation{Institute of Frontier and Interdisciplinary Science, Key Laboratory of Particle Physics and Particle Irradiation (Ministry of Education), Shandong University, Qingdao 266237, China}

\author{Gui-Rong Liang}
\email[Corresponding author: ]{bluelgr@sina.com}
\affiliation{Department of Physics, Southern University of Science and Technology, Shenzhen 518055, China}
\affiliation{School of Materials Science and Physics, China University of Mining and Technology, Xuzhou 221116, China}

\author{Zhan-Feng Mai}
\email{zhanfeng.mai@gmail.com}
\affiliation{Kavli Institute for Astronomy and Astrophysics, Peking University, Beijing 100871, China}
\affiliation{Guangxi Key Laboratory for Relativistic Astrophysics,
School of Physical Science and Technology, Guangxi University, Nanning 530004, China}

\author{Xin Zhang}
\email{zhangxin@mail.neu.edu.cn}
\affiliation{Key Laboratory of Cosmology and Astrophysics (Liaoning) \& College of Sciences, Northeastern University, Shenyang 110819, China}
\affiliation{Key Laboratory of Data Analytics and Optimization for Smart Industry (Ministry of Education), Northeastern University, Shenyang 110819, China}
\affiliation{National Frontiers Science Center for Industrial Intelligence and Systems Optimization, Northeastern University, Shenyang 110819, China}

\title{Superradiant growth anomaly magnification in evolution of vector bosonic condensates bounded by a Kerr black hole with near-horizon reflection}

\begin{abstract}
    Ultralight vector particles can form evolving condensates around a Kerr black hole (BH) due to superradiant instability. We study the effect of near-horizon reflection on the evolution of this system: by matching three pieces of asymptotic expansions of the Proca equation in Kerr metric and considering the leading order in the electric mode, we present explicit analytical expressions for the corrected \replaced{spectrum}{energy level shifts} and the superradiant instability rates.
\deleted{; the duration of whole evolution is prolonged with a factor calculated to be $(1+\mathcal{R})/({1-\mathcal{R}})$ approximately}
Particularly, in high-spin BH cases, we identify an anomalous situation where the superradiance rate is temporarily increased by the reflection parameter $\mathcal{R}$, which also occurs in the scalar scenario, but is largely magnified in vector condensates due to a faster growth rate in dominant mode. {We point out the condition for the growth anomaly in the adiabatic case is that information carried per particle exceeds a certain value $\d I/\d N>2\pi k_\text{B} \sqrt{(1+\ma R)/(1-\ma R)}$}. We further construct several featured quantities to illustrate it, and formalize the anomaly-induced gravitational wave strain deformation. 
\end{abstract}
\maketitle
\tableofcontents

\newpage
\section{Introduction}

Ultralight bosonic particles, which are promising predictions in a variety of beyond-Standard-Model theories \cite{Arvanitaki:2009fg,Essig:2013lka,Irastorza:2018dyq,Graham:2015rva,Agrawal:2018vin,Antypas:2022asj} 
have caught much attention as \deleted{one of the} competitive candidates for dark matter
\deleted{(DM)}
\cite{Preskill:1982cy,Abbott:1982af,Dine:1982ah,Svrcek:2006yi,Arvanitaki:2010sy,Brito:2015oca,Marsh:2015xka,Hui:2016ltb,Annulli:2020lyc,Chadha-Day:2021szb,Davoudiasl:2020uig,Adams:2022pbo,Li:2013nal}.
\deleted{, and are categorized as ultralight dark matter (UDM).}
{They originally arise from the Peccei-Quinn axion \cite{Peccei:1977hh} introduced to solve the strong CP problem, and later enlarged to include familions \cite{Wilczek:1982rv} and Majorons \cite{Chikashige:1980ui} from spontaneous symmetry breaking;
string theory could even suggest simultaneous presence of numerous axion-like particles taking up the mass spectrum down to the Planck mass \cite{Mehta:2020kwu}, 
filling the gap between observed massless and very massive elementary bosons. 
Effective scalar degrees of freedom can also naturally arise from several modified theories of gravity \cite{Berti:2015itd,Sotiriou:2011dz}
. In addition to scalar/pseudo-scalar fields, the zoo of light bosons extends to those with spin: massive vector fields, or known as dark photons \cite{Ackerman:2008kmp,Nakayama:2019rhg}, arise in the hidden U(1) sector \cite{Goodsell:2009xc,Jaeckel:2010ni,Camara:2011jg,Goldhaber:2008xy,Proceedings:2012ulb}; massive tensor fields, with the potential to modify the gravitational
interaction at large scale, could possibly account for the acceleration of cosmic expansion \cite{Hinterbichler:2011tt,deRham:2014zqa}
.} While traditional detections of such {ultralight 
bosonic particles in laboratory} 
are {extremely} challenging, 
\deleted{there are alternative methods that may provide effective evidence of their existence,}
{their Bose-Einstein condensates around a massive Black Hole (BH), via the mechanism of superradiance, may provide interesting phenomenology and possible observations at astrophysical and cosmological scales \cite{Hlozek:2014lca, Mehta:2020kwu}.
}

{Superradiance, which is a well-known phenomenon in flat spacetime, exhibits the amplification of incoming energy and give it back outwards. 
Its rotational case manifestly show up in the tidal heating and acceleration in the Earth-Moon system, in which frictions between the Earth and the tidal bulge would at the same time transfer the energy to heat and lifts the Moon's orbit \cite{Brito:2015oca}. Superradiance is a consequence of second law of thermodynamics, with dissipation the most essential ingredient.
In BH physics, the effect of dissipation is induced by the ergosphere, where spacetime is dragged like the tidal bulge; this ``viscous region" makes the energy extraction from BHs possible, known as the Penrose process, with its wave analog amplifies the magnitude of  incoming amplitude. The superradiance condition can be directly given by the BH area theorem, although the logic is originally reverse.
The BH first law relates its change in mass $M$, angular momentum $J$, horizon area $A$, in the rotational but neutral case as 
\be
\d M=\frac{\k}{8\pi}\d A +\O_\text{H}\d J
\ee
with $\k$ the surface gravity and $\O_\text{H}$ the horizon angular velocity.
A wave with frequency $\o$ and an azimuthal number $m$ carried away gives rise to the BH angular momentum and mass loss ratio as $\d J/\d M=m/\o$, this reshapes the above first law as
\be
\protect\label{difam}
\d A=\frac{8\pi}{\k} \frac{(\o-m\O_\text{H})}\o \d M,
\ee
joining with the second law $\d A>0$ and the mass loss in BH $\dif M<0$ requires
\be
\protect\label{spd_cond}
\o<m \O_\text{H},
\ee 
which is 
the superradiance condition. 
However, if the amplification happens only once, the back-reaction to the spacetime is verified to be at second orders of perturbation \cite{Baake:2016oku,Brito:2015oca}, i.e., the spacetime is stable. The instability occurs when a reflective boundary is put far from the BH, bouncing the outcoming waves back to the BH over and over again, which can be achieved by putting a mirror 
in toy models \cite{Press:1972zz,Cardoso:2004nk}, or in AdS spacetimes \cite{Hawking:1999dp,Cardoso:2006wa,Uchikata:2009zz}; while in realistic astrophysical contexts, the mass of the boson itself serves as a natural mirror to confine its distribution range, thus forming a quasi-bounded system, known as a ``gravitational atom" \cite{Baumann:2019eav},
featured with a
``gravitational fine structure constant". The constant is
taken from atomic physics by dividing $\hbar c$ into the coefficient of inverse square law, 
\be\protect\label{alpha}
\a\equiv \frac{GM\m}{\hbar c}
=\frac{r_g}{\l_c}=\frac{\l_c}{r_B}
\simeq 0.02\(\frac M{3M_\odot}\)\(\frac\m{10^{-12}\text{eV}}\),
\ee
where $\m$ is the boson mass
considered to be a free parameter, with $\l_c=\hbar/\m c$ the corresponding Compton wave length, $r_g=GM/c^2$ and  $r_B=\hbar^2/(GM\m^2)$ are the gravitational radius and the Bohr radius respectively.
 In such an atomic structure, the BH-boson system exhibits a hydrogen-like spectrum but with growing occupation numbers at each energy level. Predicted monochromatic gravitational waves (GW) are generated either from transitions or annihilations between or within these levels for a long period of time \cite{Damour:1976kh,
Zouros:1979iw,Detweiler:1980uk,Dolan:2007mj,Shlapentokh-Rothman:2013ysa,Pani:2012vp,Pani:2012bp,Witek:2012tr,Brito:2013wya,East:2017mrj,East:2017ovw,Baryakhtar:2017ngi,Cardoso:2018tly,East:2018glu,Frolov:2018ezx,Dolan:2018dqv,Baumann:2019eav,Brito:2020lup,Davoudiasl:2021ijv,Chung:2021roh,PPTA:2022eul,Yuan:2022bem}, potentially to put constraints on the physical parameters of these ultralight bosonic dark matters \cite{Arvanitaki:2014wva,Brito:2017wnc,Brito:2017zvb,Zhu:2020tht,Kodama:2011zc,Brito:2014wla,Tsukada:2018mbp}. 
}

The growing condensate is characterized by the superradiance rate, 
for a bosonic state with spin-$s$, and in the limit of small mass coupling $\a$, the rate \cite{Baryakhtar:2017ngi,Brito:2020lup} scales as 
\be\label{g_bhav}
\G_{nljm}\propto(m\O_\text{H}-\o)~\a^{2(l+j)+5},
\ee
with the orbital angular momentum $l$ and the total angular momentum number $j\in(|l-s|,l+s)>0$ integers, and $m\in(-j,j)$ the quantum projection of $j$ along the rotation axis. The spins $s=0,1,2$ are for scalar, vector, and tensor respectively. The superradiance condition~\eqref{spd_cond} guarantees the positivity of the rate, otherwise the state becomes absorptive, which occurs definitely if $m=0$ or possibly if bosons are too heavy bringing a large $\o$; while too light bosons would lead to a much slower growth due to the $\a$-suppression.
The optimal case occurs when $\o\simeq 
\O_\text{H}$ and a BH is near-extremal, this
gives $\a\simeq 0.5$ or equivalently $\l_c\simeq 2r_g$,
which says when the bosonic Compton wavelength is around the radius of BH, the superradiance becomes most significant.
In terms of different states, the largest rate occurs 
 at smallest $(l+j)$ but maximum quantum number $m=j$, which will consequently lead to a larger GW amplitude during the superradiance.  In Table~\ref{modes}, we list several leading modes for scalar, vector and tensor field, according to different orbital angular momentum number $l$, for largest total angular momentum $j=l+s$ and largest quantum number $m=j$
 ; and in Figure~\ref{svt_figs}, we plot schematic pictures of dominant scalar, vector and tensor modes around a spinning BH.\\
~\\
\begin{minipage}{0.45\textwidth}\label{modes}
\vspace{42pt}
{\centering
\resizebox{0.8\textwidth}{!}{
	\setlength{\tabcolsep}{0.1pt}{
	\begin{tabular}{|l|c|c|c|}
	\hline
	&l=0&l=1&l=2\\
	\hline
	Scalar: $|nlm\>_{(j=l)}$ &&\cellcolor{orange!70}{{$|211\>$}}&$|322\>$\\
	\hline
	Vector: $|nljm\>$ &\cellcolor{red!70}{$|1011\>$}& $|2122\>$&$|3233\>$\\
	\hline
	Tensor: $|nljm\>$&\cellcolor{orange!70}$|1022\>$&$|2133\>$&$|3244\>$\\
	\hline
	\end{tabular}}}\\
\vspace{30pt}}
{\small TABLE I. Leading superradiant states of different bosonic fields in terms of orbital angular momentum number $l$, for largest total angular momentum $j=l+s$ and largest quantum number $m=j$. The dominant modes are filled with colors, with the superior vector case $|1011\>$ in red, the following $|211\>$ and $|1022\>$ for scalar and tensor modes in orange, parallel to the right schematic pictures.}
\vspace{10pt}
\end{minipage}
\begin{minipage}{0.2\textwidth}
\end{minipage}
\begin{minipage}{0.5\textwidth}\label{svt_figs}
\vspace{-22pt}
{\centering
\begin{tabular}{c}
\adjustbox{valign=t,scale=0.1}{
               \begin{tikzpicture}
    \def\radius{6};
    \def\arrowlength{1};

    \phantom{\shade[inner color={rgb:red,255;green,42;blue,0}] (0,0) circle (\radius);}

    \def\numa{6pt};
    \def\numb{8pt};
    \foreach \i in {0,1,...,\numa-1} {
        \phantom{\pgfmathsetmacro{\x}{0.5*\radius*cos(\i*360/\numa)}
        \pgfmathsetmacro{\y}{0.5*\radius*sin(\i*360/\numa)}
         \draw [-Stealth, line width=1mm] (\x,\y-0.5*\arrowlength) -- (\x,\y+0.5*\arrowlength);}}

        \foreach \i in {0,1,...,\numb-1} {
        \phantom{\pgfmathsetmacro{\x}{0.85*\radius*cos(\i*360/\numb)}
        \pgfmathsetmacro{\y}{0.85*\radius*sin(\i*360/\numb)}
       \draw [-Stealth, line width=1mm] (\x,\y-0.5*\arrowlength) -- (\x,\y+0.5*\arrowlength);}}

    \phantom{\draw [-Stealth, line width=1.5mm, color=cyan] (0,-2) -- (0,+2);}
    \phantom{\shade[ball color=black] (0,0) circle (1);}
\end{tikzpicture}
    
    }
    \adjustbox{valign=t,scale=0.35}{
               \begin{tikzpicture}
  \def\length{4}
  \def\width{6}
  \def\edge{0}

  \shade[inner color=orange] 
    (\edge,0) 
    .. controls (\edge+\width/2, -\length/2) and (\edge+\width, -\length/2) .. (\edge+\width, 0)
    .. controls (\edge+\width, \length/2) and (\edge+\width/2, \length/2) .. (\edge,0)
    -- cycle;

  \shade[inner color=orange] 
    (-\edge,0) 
    .. controls (-\edge-\width/2, -\length/2) and (-\edge-\width, -\length/2) .. (-\edge-\width, 0)
    .. controls (-\edge-\width, \length/2) and (-\edge-\width/2, \length/2) .. (-\edge,0)
    -- cycle;

    \draw [-Stealth, 
    line width=1.5mm, 
    color=cyan] (0,-2) -- (0,+2);
    \shade[ball color=black] (0,0) circle (1);
	
\end{tikzpicture}
    }\\
    
    \adjustbox{valign=c,scale=0.2}{
               \begin{tikzpicture}
    \def\radius{6};
    \def\arrowlength{1};

    \shade[inner color={rgb:red,255;green,42;blue,0}] (0,0) circle (\radius);

    \def\numa{6pt};
    \def\numb{8pt};
    \foreach \i in {0,1,...,\numa-1} {
        \pgfmathsetmacro{\x}{0.5*\radius*cos(\i*360/\numa)}
        \pgfmathsetmacro{\y}{0.5*\radius*sin(\i*360/\numa)}
        \draw [-Stealth, line width=1mm] (\x,\y-0.5*\arrowlength) -- (\x,\y+0.5*\arrowlength);}

        \foreach \i in {0,1,...,\numb-1} {
        \pgfmathsetmacro{\x}{0.85*\radius*cos(\i*360/\numb)}
        \pgfmathsetmacro{\y}{0.85*\radius*sin(\i*360/\numb)}
        \draw [-Stealth, line width=1mm] (\x,\y-0.5*\arrowlength) -- (\x,\y+0.5*\arrowlength);}

    \draw [-Stealth, line width=1.5mm, color=cyan] (0,-2) -- (0,+2);
    \shade[ball color=black] (0,0) circle (1);
\end{tikzpicture}
    }
    \adjustbox{valign=c,scale=0.4}{
               \begin{tikzpicture}
  \def\length{4}
  \def\width{6}
  \def\edge{0}

  \phantom{\shade[inner color=orange] 
    (\edge,0) 
    .. controls (\edge+\width/2, -\length/2) and (\edge+\width, -\length/2) .. (\edge+\width, 0)
    .. controls (\edge+\width, \length/2) and (\edge+\width/2, \length/2) .. (\edge,0)
    -- cycle;}

  \phantom{\shade[inner color=orange] 
    (-\edge,0) 
    .. controls (-\edge-\width/2, -\length/2) and (-\edge-\width, -\length/2) .. (-\edge-\width, 0)
    .. controls (-\edge-\width, \length/2) and (-\edge-\width/2, \length/2) .. (-\edge,0)
    -- cycle;}

    \phantom{\draw [-Stealth, 
    line width=1.5mm, 
    color=cyan] (0,-2) -- (0,+2);
    \shade[ball color=black] (0,0) circle (1);}

\end{tikzpicture}
    }\\
    
    \adjustbox{valign=b,scale=0.2}{
                \begin{tikzpicture}
    \def\radius{6};
    \def\arrowlength{1.5};

    \shade[inner color=orange] (0,0) circle (\radius);

    \def\numa{6pt};
    \def\numb{8pt};
    \foreach \i in {0,1,...,\numa-1} {
        \pgfmathsetmacro{\x}{0.5*\radius*cos(\i*360/\numa)}
        \pgfmathsetmacro{\y}{0.5*\radius*sin(\i*360/\numa)}
        \draw [-Stealth, line width=1mm] (\x,\y-0.5*\arrowlength) -- (\x,\y+0.5*\arrowlength);}

        \foreach \i in {0,1,...,\numb-1} {
        \pgfmathsetmacro{\x}{0.85*\radius*cos(\i*360/\numb)}
        \pgfmathsetmacro{\y}{0.85*\radius*sin(\i*360/\numb)}
        \draw [-Stealth, line width=1mm] (\x,\y-0.5*\arrowlength) -- (\x,\y+0.5*\arrowlength);}

    \draw [-Stealth, line width=1.5mm, color=cyan] (0,-2) -- (0,+2);
    \shade[ball color=black] (0,0) circle (1);
\end{tikzpicture}
    }
    \adjustbox{valign=c,scale=0.4}{
               \begin{tikzpicture}
  \def\length{4}
  \def\width{6}
  \def\edge{0}

  \phantom{\shade[inner color=orange] 
    (\edge,0) 
    .. controls (\edge+\width/2, -\length/2) and (\edge+\width, -\length/2) .. (\edge+\width, 0)
    .. controls (\edge+\width, \length/2) and (\edge+\width/2, \length/2) .. (\edge,0)
    -- cycle;}

  \phantom{\shade[inner color=orange] 
    (-\edge,0) 
    .. controls (-\edge-\width/2, -\length/2) and (-\edge-\width, -\length/2) .. (-\edge-\width, 0)
    .. controls (-\edge-\width, \length/2) and (-\edge-\width/2, \length/2) .. (-\edge,0)
    -- cycle;}

    \phantom{\draw [-Stealth, 
    line width=1.5mm, 
    color=cyan] (0,-2) -- (0,+2);
    \shade[ball color=black] (0,0) circle (1);}

\end{tikzpicture}
    }
\end{tabular}\\}
\vspace{-15pt}
{\small FIG. I. Schematic pictures of dominant scalar, vector and tensor modes around a spinning BH. The BH spins are denoted by blue arrows, the boson spins are denoted by black arrows; the superior vector condensate $|1011\>$ is in red, the following $|211\>$ and $|1022\>$ for scalar and tensor clouds are in orange, parallel to the left table fillings. The distribution for scalar is for $l=1$, while those for vector and tensor are for $l=0$; the arrows for tensors are $2$ times as those of vectors. }
\vspace{10pt}
\end{minipage}\\


{It is clearly seen that the vector state $|1011\>$ with $l+j=1$ has the largest growing rate, while both the scalar state $|211\>$ and the tensor state $|1022\>$ with $l+j=2$ are $\a^2$ orders of magnitude inferior\footnote{{The only exception is the tensor polar dipole mode \cite{Brito:2013wya}, growing uniquely with a non-hydrogenic form as $\G\propto (m\O_\text{H}-\o)~\a^3$, which is beyond our current research scope.}}. {Further, a special feature of superradiantly-generated clouds with a vanishing orbital angular momentum dominant growing mode, say $|1011\>$ or $|1022\>$, is that the distribution of the cloud is much closer to the BH horizon and the BH rotation energy is solely transferred into the bosonic intrinsic spin \cite{Baumann:2019eav}; these makes the states 
especially sensitive to the near-horizon geometry of the BH. For these reasons, we highly suggest that the vector dominant state $|1011\>$ to serve as the best candidate to investigate phenomenology induced by near-horizon geometry, which is our main focus in this paper, see Fig. I. for a schematic picture.}}

{The BH near-horizon geometry is trivial in classical theory of {particle motions in} general relativity, in which an event horizon is treated as one-way membrane where nothing can escape,}
superradiance in this case has been studied in different bosonic fields \cite{Bao:2022hew,Bao:2023xna,Guo:2024dqd,Ficarra:2018rfu,Guo:2022mpr}.
While the BH information paradox \cite{Polchinski:2016hrw} suggests \deleted{the} modifications near the BH horizon, which could be given in some {semi-classical treatments of particle motions \cite{Kuchiev:2003ez,Kuchiev:2003rf,Kuchiev:2003rv,Kuchiev:2003fy,Kuchiev:2004fn,Flambaum:2004ru,Kuchiev:2005em}, or several} quantum gravity theories \deleted{also predict quantum-corrected BH models} \cite{Skenderis:2008qn,Haggard:2014rza,Bianchi:2018mml} .
The quantum-corrected models\deleted{, in general,} often involve {these} modifications \deleted{at the BH horizon} by varying the boundary condition when a test field presents \deleted{around the BH}, with a reflection parameter $\ma R$ introduced as an effective phenomenological description.
\deleted{There is an effective phenomenological description of these modifications by introducing a reflection parameter $\mathcal{R}$.} \replaced{This method}{ The reflective feature} is commonly 
\replaced{employed}{presented} in the study of GW echoes \cite{Cardoso:2016rao,Cardoso:2019apo,Mark:2017dnq,Price:2017cjr,Fang:2021iyf,Biswas:2022wah,Bueno:2017hyj,Ikeda:2021uvc,Abedi:2016hgu}, 
{and has been introduced to investigate possible effects on superradiance in scalar case in recent years \cite{Guo:2021xao,Guo:2023mel,Zhou:2023sps,Luo:2024gqo}. But as we suggested earlier, the vector dominant state should be the best candidate to study near-horizon geometry, which is our focus in this paper. We will mainly investigate the influence of a real-valued reflection parameter $\ma R$ on the superradiance rate $\G(\ma R)$ and draw possible conclusions on the evolution and the corresponding GW patterns.
}

The
 paper is organized as follows. In Section~\ref{sec:setup}, we employ the Proca field to describe the condensate around a Kerr BH; using a non-relativistic approximation, we obtain the hydrogenic spectrum of the vector field at a large distance from the BH,
\replaced{with a complexified frequency to include the superradiant mode}{ which provides a general view} of the quasibound BH-condensate system. 
In the leading-order of the angular part, the method of matched asymptotic expansions in radial part, in three pieces in our context, is used to obtain 
the analytical results. {We optimize the matchings with near region solution by choosing a different basis, which provide simpler but natural connections to the boundary condition.}
In Section~\ref{sec:nearhorizon}, we introduce the reflection parameter describing a modified boundary condition at the horizon, and \replaced{investigate}{calculate} its modifications to \deleted{the} energy levels and {the} superradiance rate analytically. Particularly, in high-spin BH and low-frequency particle cases,  the superradiance rate may  temporally get boosted {by the reflection}, rather than monotonically being slowed down. This also occurs in scalar scenario, but is largely magnified in vector case due to the faster growth rate and closer distribution geometry, as shown in Table~\ref{modes} and in Figure~\ref{svt_figs}, and is named as an ``anomaly" in our context. {We give three forms of the anomaly condition, which have illustrative, geometrical, and physical meanings respectively, and point out this is due to a ``rotation-relativistic" effect of fast-spinning BHs.}
In Section~\ref{sec:evo}, we evaluate the influence of the reflection parameter on the whole system's evolution, finding that the above ``anomaly" only lasts temporally at the early stage, without affecting the final states too much, but it is most obvious in the vector case due to a overwhelming faster growth rate in dominant mode. This scenario is much like a version of ``The Tortoise and the Hare",
{with ``the Tortoise" the normal decreasing superradiant rate with reflection,} and ``the Hare" the temporal anomalous superradiant boost. 
We introduce three featured quantities to describe the phenomenon:
the maximum growth rate difference, 
the maximum condensate mass difference,
and the advantage time of the anomaly.
Further, we formalize the anomaly and estimate
\deleted{with the ``magnification factor" $\mathcal{F}_{\text {mag }}$ and 
a ``spin deviation factor" $\mathcal{F}_{\mathrm{dev}}$}
its effects on GW strain in the early stage. 
In Section~\ref{sec:conclusion}, we conclude our main procedure and results, followed by which {a possible physical explanation for the anomaly is proposed}, discussions and some extensions of the study are presented in the end. Throughout this paper, we adopt the Planck units $G_\mathrm{N}=\hbar=c=1$ and the $(+,-,-,-)$ signature, {and a small coupling $\a=M\m\ll1$ approximation in the analytical analysis.}

\section{Hydrogenic spectrum for vector bosonic condensates and Proca equation in Kerr spacetime}
\protect\label{sec:setup}
In this section, we study the vector bosonic condensate around a Kerr black hole by solving the Proca equation in a Kerr metric; we will firstly demonstrate its non-relativistic approximate solutions in distant region from the horizon which exhibits hydrogenic features, then we will elaborate the full analytical treatment of matched asymptotics for the radial part, by considering the leading order of electric mode in the separated angular function. 

The vector condensate is described by a free massive spin-1 field $A_\mu$, the corresponding Lagrangian in an arbitrary spacetime is given by
\begin{equation}
\mathcal{L}_{\mathcal{A}}=-\frac{1}{4}F_{\rho\sigma}F^{\rho\sigma}+\frac{1}{2}\mu^2A_{\gamma}A^{\gamma},
\end{equation}
with $\mu$ the particle mass and the associated field tensor as
\begin{equation}\protect\label{eq:field_strength_tens}
F_{\rho\sigma} \equiv \nabla_\rho A_\sigma-\nabla_\sigma A_\rho,
\end{equation}
where indices are raised and lowered by the metric $g_{\rho\sigma}$. 

Taking the Euler-Lagrangian equation for $A^{\mu}$, the resulting Proca equation of motion (EoM) is
\begin{equation}
    \nabla_\rho F^{\rho \sigma}+\mu^2 A^\sigma=0,
    \protect\label{eq:Proca}
\end{equation}
with the Lorenz gauge condition $\nabla_\rho A^\rho=0$ automatically satisfied; the EoM and gauge condition together constrain components of $A^{\mu}$ to have $3$ independent degrees of freedom, which we choose to be the spatial components afterwards. The information of energy and momentum density and flux is encoded in the corresponding stress-energy tensor, which is given by
\begin{equation}
\begin{aligned}
    T_{\rho\sigma}=&F_{\rho\lambda}F_{\sigma}{}^{\lambda}-\mu^2A_{\rho}A_{\sigma}
    +g_{\rho\sigma}\left(-\frac{1}{4}F_{\alpha\beta}F^{\alpha\beta}+\frac{1}{2}\mu^2A_{\gamma}A^{\gamma}\right).
\end{aligned}    
\end{equation}

In the superradiance regime, the black hole rotation triggers the bosonic condensate cloud growth; the most common astrophysical rotating Kerr BHs are dressed with only two distinguished features: the mass $M$ and the angular momentum $J$, the corresponding metric in Boyer-Lindquist coordinates takes the form
\begin{equation}
\begin{aligned}
\mathrm{d}s^2=\left( 1-\frac{2Mr}{\Sigma} \right) \mathrm{d}t^2&+\frac{4aMr}{\Sigma}\sin ^2\theta~ \mathrm{d}t\mathrm{d}\varphi -\frac{\Sigma}{\Delta}\mathrm{d}r^2-\Sigma \mathrm{d}\theta ^2
\\
&-\left[ \left( r^2+a^2 \right) \sin ^2\theta +2\frac{Mr}{\Sigma}a^2\sin ^4\theta \right] \mathrm{d}\varphi ^2,
\protect\label{eq:KerrMetric}
\end{aligned}
\end{equation}
with
\begin{equation}
\left\{\begin{aligned}
\Delta&\equiv r^2-2 M r+a^2,\\
\Sigma&\equiv r^2+a^2 \cos ^2 \theta,
\end{aligned} \right.  
\end{equation}
and $a\equiv J/M$ the BH angular momentum per unit mass.
We will also refer to the dimensionless spin parameter $\chi\equiv a/M= J/M^2$ as the ``BH spin" in the following context.

The inner and outer horizons $r_\pm$ and infinite redshift surfaces $r^s_\pm$ of the Kerr BH are given by
\begin{equation}
\left\{\begin{split}
r_{ \pm}&=M \pm \sqrt{M^2-a^2}\\
r^s_{ \pm}&=M \pm \sqrt{M^2-a^2\cos^2\theta}.
\end{split}\right.
\end{equation}
The classical analogue of superradiance, Penrose process, happens in ergosphere between $r_+$ and $r_+^s$ where negative energy trajectories exist; the different boundary conditions with or without reflection are chosen at the outer horizon $r_+$, since the inner horizon $r_-$ is not physically accessible. The angular velocity of the outer horizon is given by $\O_\text{H}\equiv a/2Mr_+=\chi/2r_+$.

{Note that the quasi-adiabatic approximation is assumed when we employ the stationary Kerr metric \eqref{eq:KerrMetric}. The bosonic field is typically distributed over a very large volume, implying very small densi- ties and consequent small backreaction effects. It is treated as perturbation on the Kerr BH, giving a dynamical contribution to the background with a very small timescale, compared to typical oscillation period of the boson field and the instability timescale \cite{Brito:2014wla}. In other words, the BH is soon going back to the stationary case after the perturbation in the previous time instant, before we can use it as a background to continue the superradiance at the next instant.
}

\subsection{Hydrogenic solutions in non-relativistic approximation and spectrums}
As a first step, we will present 
a preliminary investigation into the solutions using a non-relativistic approximation, since
its fundamental structure would
 exhibit key features of the gravitational atom by highly similar analytical energy levels, which could be sources of monochromatic GWs with waveform potentially to be influenced by superradiance with reflections.

We consider the situation where the Compton wavelength ${\lambda}_\text{Compton}\sim\mu^{-1}$ of the vector particles is substantially larger than the gravitational radius $r_g\equiv M$ of the BH, or equivalently, 
the mass coupling of the BH-condensate system $\alpha\equiv M\mu$ is much less than 1, 
{hence
our considered region is restricted to be located distantly from the BH horizon.} Under these conditions, the solutions to the Proca equation in the Kerr metric will exhibit nonrelativistic and hydrogen-like properties \cite{Baumann:2019eav}. Hence, a single solution can be expressed in the following form \cite{Baryakhtar:2017ngi}:
\begin{equation}
  A^{\r}(t, x)=\frac{1}{\sqrt{2 \mu}}\left(\Psi^{\r}(x) \mathrm{e}^{-\mathrm{i} \omega t}+\text { c.c. }\right) .
  \protect\label{eq:vector_function_Single}
\end{equation}

\deleted{Given the small mass coupling of the system, we can assume that the condensate is mainly far from the BH. }
For $r \gg r_g$, substituting Eq.~\eqref{eq:vector_function_Single} into the Proca equation \eqref{eq:Proca}, and approximating the metric tensor so that it has flat space-space components $g_{ij}=\eta_{ij}$ and vanishing off-diagonal components $g_{0i}=0$, and only a nontrivial component $g_{00}$, we obtain the following approximate EoM:
\begin{equation}
\left(\omega^2-\mu^2\right) \Psi^\r \approx -\nabla^2 \Psi^\r-\omega^2\left(-1+g^{00}\right) \Psi^\r,
\end{equation}
with the component $g^{00} \approx\left(1- 2r_{g} / r\right)$ in the Kerr metric at $r \gg r_g$. The spatial change in $\Psi^\mu$ is gradual, resulting in non-relativistic momenta of the particles, so the frequency of the bound state is approximately equal to its mass, $\omega \approx \mu$, to the leading order in $\alpha$; thus the EoM further simplifies to 
a Schrödinger-type equation with a potential of $1/r$:
\begin{equation}
(\omega-\mu) \Psi^{\r} \approx -\frac{\nabla^{2}}{2 \mu} \Psi^{\r}+\frac{\alpha}{r} \Psi^{\r}.
\end{equation}

As stated above, we choose the $3$ independent degrees of freedom to be the spatial components, then the temporal component is determined by the Lorentz gauge condition.
Given the spherically symmetric feature of the $1/r$ potential, 
the spatial components can be solved by separation of variables:
\begin{equation}
\Psi_i=R^{n l}(r) Y_i^{l, j m}(\theta, \varphi),
\end{equation}
with $\{n,l,j,m\}$ the principle, orbital angular momentum, total angular momentum, and azimuthal angular momentum numbers respectively, and totally determine the vector eigenstates; note that the scalar eigenstates are determined only by three numbers $\{n,l,m\}$, since there're no intrinsic spin of the scalar cloud itself. The orbital angular momentum number $l$ is taken to be from $0$ to $n$, and denoted by $\{$s, p, d, f...$\}$.
The ``pure-orbital vector spherical harmonics"
 $Y_i^{l, j m}(\theta, \varphi)$ are eigenfunctions of the orbital angular momentum operator \cite{Thorne:1980ru,Baryakhtar:2017ngi}, satisfying
\begin{equation}
    -r^{2} \nabla^{2} Y_{i}^{l, j m}=l(l+1) Y_{i}^{l, j m},
\end{equation}
with $j$ taken to be $j=l\pm 1, l$, due to the spin-$1$ nature of vector field. The $3$ degrees of freedom of the vector field can thus be transferred to be this $3$ values of the total angular momentum number; the $j=l\pm 1$ modes are known as the ``electric modes" and the $j=l$ modes as the ``magnetic modes", classified by their parity features \cite{Thorne:1980ru, Baumann:2019eav}. In the later discussion of more rigorous treatment, we will only focus on the electric modes, due to its mathematical convenience and exemplariness.

The radial solutions $R^{n l}(r)$ are hydrogenic wavefunctions, which can be expressed in terms of the associated Laguerre polynomials $L_{n}^{2l+1}(r)$ as in \cite{Brito:2014wla},
\begin{equation}\protect\label{hr}
    R_{nl}(\tilde{r})=\sqrt{\left(\frac{2}{\bar{n}r_{a}}\right)^3\frac{n!}{2\bar{n}(\bar{n}+l)!}}\mathrm{e}^{-\tilde{r}/2}\tilde{r}^l L_{n}^{2l+1}(\tilde{r}),
\end{equation}
where we changed our variable to be $\tilde{r}\equiv 2r/(\bar{n}r_a)$, with $r_a=1/(M\mu^2)$ the ``Bohr radius" and principle quantum number $\bar{n}= n+l+1$ labeling the ``energy levels" of the system, analogous to their quantum mechanical counterparts. Thus we stated the non-relativistic approximation exhibits hydrogenic features in the beginning of this subsection; more explicitly, 
these levels are presented in a highly similar form with that of hydrogen atom 
\begin{equation}
    \omega\approx\mu\left(1-\frac{\alpha^2}{2\bar{n}^2}\right),
    \protect\label{eq:Approxomega}
\end{equation}
{where a higher order correction in $\alpha$ is added, to the leading order approximation $\omega\approx\mu$.}
Identification of these discrete energy levels allows two channels for monochromatic GW emission, from level transitions or annihilations \cite{Arvanitaki:2014wva}, with angular frequencies
\be\protect\label{2cl}
\1\{\begin{split}
\o_{\text{tr}}&=\frac12\m \a^2\(\frac1{\bar n_\text{H}^2}-\frac1{\bar n_l^2}\)\\
\o_{\text{ann}}&=2\o\simeq 2\m\(1-\frac{\a^2}{2\bar n^2}\),
\end{split}\2.
\ee
with $\bar n_\text{H}$ and $\bar n_l$ principle quantum numbers of higher and lower energy levels.
\deleted{Thus any possible reason that shifts these energy levels, e.g, reflection due to the near-horizon boundary condition in our contexts, would lead to deviations from monochromaticity and hence influence the waveforms of this type.}

{
More generally, in case of rotation, superradiant instability is given by a complex shift to the frequency of this Bohr-like spectrum as 
\be\protect\label{nnn}
\o = \omega _0+\delta \omega\equiv \m \sqrt{1-\frac{\a^2}{\n^2}},
\ee
where
the shifted and complexified energy level is defined as $\nu \equiv\bar{n}+\delta \nu$ with $\delta \nu \in \mathbb{C} ,|\delta \nu |\ll \bar n$, then
the two parts of the spectrum are extracted as
\begin{equation}\protect\label{odo}
    \1\{\begin{aligned}
        \omega_0 &=\mu \left( 1-\frac{\alpha ^2}{2\bar{n}^2} \right) ,
        \\
        \delta \omega &=\frac{M^2\mu ^3}{\bar{n}^3}\delta \nu.
    \end{aligned}\2.
\end{equation}
we see that $\omega_0$ is just the Bohr-like spectrum in \eqref{eq:Approxomega}, and the complex shift $\delta\omega$ is controlled by the shift in energy levels $\delta\n$. If we write the spectrum in explicit real and imaginary part as $\omega =E+\mi \G$, with $E$ the new energy spectrums and $\G$ the superradiant instability rate,
we have
\begin{equation}\protect\label{EG}
    \1\{\begin{aligned}
       E&=\omega _0+\mathrm{Re}\left( \delta \omega \right)
\\
\Gamma &=\mathrm{Im}\left( \delta \omega \right),
    \end{aligned}\2.
\end{equation}
which is solely determined by $\del\o$ and hence by $\del\n$.
Thus the information of both spectrum and superradiant growth is totally encoded in the complex energy level $\n$, 
containing any possible changes by various potential modifications. In the following contents, the main task is to determine $\del\n$ in case of boundary conditions with the near-horizon reflection parameter $\ma R$, and to discuss the effects on superradiant growth and BH-bosonic bound system evolution. Notably, the effect of reflection can be factor out as
\be\protect\label{dnn}
\d\n(\ma R)=\d\bar\n~ \Phi(\ma R)
\ee
with $\Phi(\ma R=0)=1$, thus $\d\bar\n$ denotes the modification purely due to superradiant instability without any reflection, and later we will see that $\d\bar\n$ is a pure imaginary number hence so is $\d\o$; therefore, in case of no reflection, the energy spectrum is not changed, $E=\o_0$, according to \eqref{EG}; even in case of reflection, if $\text{Re}(\d\o) \ll \o_0$, we can treat the energy spectrum as unchanged, and focus on the superradiance part, which will be our case in the later analysis.
}

\subsection{Analytical description with separable ansatz and leading order behaviors}

The non-relativistic approximation provides a general energy spetrum of the Proca condensate, but by dropping the off-diagonal metric components, it lost the key ingredient of Kerr geometry --- the rotation, which is essential to trigger superradiance. To formally treat the problem, 
one has to solve the equation by substituting the Proca equation \eqref{eq:Proca} into the Kerr metric \eqref{eq:KerrMetric}. However, 
it is the rotation that complicates the analytical resolution, since the Proca equation in the Kerr background seemingly presents an issue of separability \cite{Brito:2015oca}; several attempts were made to tackle the problem, including semi-analytical treatments \cite{Pani:2012bp,Pani:2012vp} or numerical efforts \cite{Witek:2012tr,Cardoso:2018tly,Baumann:2019eav};
afterwards, by generalizing the tools in treating Maxwell equations \cite{Lunin:2017drx,Krtous:2018bvk}, a proper  ansatz  \cite{Frolov:2018ezx} was proposed, \deleted{allowing the separation of the Proca equation in the Kerr metric}
which we will adopt in our following treatments.


The ansatz that captures all of the electric modes of a vector field on Kerr background takes the form \cite{Frolov:2018ezx,Baumann:2019eav}
\begin{equation}\protect\label{sep_ansatz}
A^\mu=B^{\mu \nu} \nabla_\nu Z, \quad\text{with}\quad
Z(t, \mathbf{r})=\mathrm{e}^{-\mathrm{i} \omega t+\mathrm{i} m \varphi} R(r) S(\theta),
\end{equation}
with $R(r)$ and $S(\th)$ the ``radial function" and the ``angular function" in terminology, respectively; their relations to the actual vector field configuration in the far zone is given in \cite{Baumann:2019eav}. The polarization tensor $B^{\mu \nu}$ is defined through the equation
\begin{equation}
B^{\mu \nu}\left(g_{\nu \sigma}+\mathrm{i} \lambda^{-1} h_{\nu \sigma}\right)=\delta_\sigma^\mu,
\end{equation}
with $\lambda$ generally a complex number, which will be referred to as the ``angular eigenvalue"; and $h_{\mu \nu}$ is called the principal tensor \cite{Frolov:2017kze} which generates the symmetries of the Kerr spacetime, with its explicit expression given in \cite{Baumann:2019eav}. For the magnetic modes, an educated hypothesis \cite{Baumann:2019eav} suggests the same energy level and superradiant instability rate of the vector field as those of the electric modes, thus we will focus only on electric modes in this paper.

With the separable ansatz, the differential equations that radial and angular functions satisfy are extracted.
The angular equation takes the form
\begin{equation}
	\frac{1}{\sin \theta}\frac{\mathrm{d}}{\mathrm{d}\theta}\left( \sin \theta \frac{\mathrm{d}S}{\mathrm{d}\theta} \right) +\left[ \Lambda -\frac{m^2}{\sin ^2\theta}+\mu ^{-2}\alpha ^2{\chi}^2\left( \omega ^2-\mu ^2 \right) \cos ^2\theta \right] S
	=\frac{2\mu ^{-2}\alpha ^2{\chi}^2\cos \theta}{\lambda ^2q_{\theta}}\left( \sin \theta \frac{\mathrm{d}}{\mathrm{d}\theta}+\sigma \lambda \cos \theta \right) S,\\   
\protect\label{eq:angularfull}
\end{equation}  
with 
\begin{equation}\1\{
\begin{aligned}
	q_{\theta}&\equiv 1-\mu ^{-2}\alpha ^2{\chi}^2\lambda ^{-2}\cos ^2\theta ,\\
	\sigma &\equiv \omega +\mu ^{-1}\alpha \chi\lambda ^{-2}(m-\mu ^{-1}\alpha \chi\omega ),\\
	\Lambda &\equiv \mu ^2\lambda ^2-\sigma \lambda +2\mu ^{-1}\alpha \chi\omega m-\mu ^{-2}\alpha ^2{\chi}^2\omega ^2.\\
\end{aligned}\2. 
\end{equation}
Note that we keep the vector mass $\m$ in the expression, which causes slight difference in expressions with those in \cite{Baumann:2019eav}.
The radial equation reads
\begin{equation}
\begin{aligned}
\frac{\mathrm{d}^2R}{\mathrm{d}r^2}+&\left( \frac{1}{r-r_+}+\frac{1}{r-r_-}-\frac{1}{r-\hat{r}_+}-\frac{1}{r-\hat{r}_-} \right) \frac{\mathrm{d}R}{\mathrm{d}r}
+\left[ -\frac{\Lambda}{\Delta}-\left( \mu ^2-\omega ^2 \right) +\frac{P_{+}^{2}}{\left( r-r_+ \right) ^2} \right. +\frac{P_{-}^{2}}{\left( r-r_- \right) ^2}
\\
&-\frac{A_+}{\left( r_+-r_- \right) \left( r-r_+ \right)}+\frac{A_-}{\left( r_+-r_- \right) \left( r-r_- \right)}
\left. -\frac{\lambda \sigma r}{\Delta \left( r-\hat{r}_+ \right)}-\frac{\lambda \sigma r}{\Delta \left( r-\hat{r}_- \right)} \right] R=0,
\end{aligned}
\protect\label{eq:originalR}
\end{equation}
with $\hat{r}_{ \pm} \equiv \pm \mathrm{i} \lambda$ depending on the angular eigenvalue, and
\begin{equation}\1\{
\begin{aligned}
\gamma ^2&\equiv \frac{1}{4}\left( r_+-r_- \right) ^2\left( \mu ^2-\omega ^2 \right) ,
\\
\gamma _{\pm}^{2}&\equiv  M^2\left( \mu ^2-7\omega ^2 \right) \pm M\left( r_+-r_- \right) \left( \mu ^2-2\omega ^2 \right),
\\
P_{\pm}&\equiv \frac{am-2M\omega r_{\pm}}{r_+-r_-},
\\
A_{\pm}&\equiv P_{+}^{2}+P_{-}^{2}+\gamma ^2+\gamma _{\pm}^{2}.
\end{aligned}\2.    
\end{equation}
The poles at  $r=r_\pm$ are the same as that in scalar radial functions which can be solved by matching near and far region solutions, while the additional poles at $r=\hat r_\pm$ appear here in vector case makes the widths of the two regions narrower thus don't overlap, thus it's necessary to introduce an intermediate region to fill the gap, which we will do in detail in the following subsection. 

{It's useful to note that the superradiance condition $\o<m \O_\text{H}$ guarantees the positivity of $P_+$, which is re-organized as
\be\protect\label{pplus}
P_+=\frac{2Mr_+}{r_+-r_-}\(m\O_\text{H}-\o\)=\frac{m\chi}{2\sqrt{1-\chi^2}}\(1-\frac\o{m\O_\text{H}}\)>0,
\ee
making the ingoing and outgoing terms $z^{\pm\mi P_+}$ explicit, with $z$ a rescaled radial coordinate in the following contexts; and in the second equator we further wrote it in terms of the BH spin $\chi$ and the ``superradiance saturation" $\o/m\O_\text{H}$,
for later convenience. Also note that in the small $\a$ limit, $P_+=P_-$.}

To solve the coupled differential equations \eqref{eq:angularfull} and \eqref{eq:originalR} systematically, expansions should be done by orders of the coupling constant $\a$. But still, hydrogenic properties exhibit in the leading order and in the far zone. The expressions with higher-order corrections can be found in Ref.~\cite{Baumann:2019eav}. 
For the electric modes, by dropping explicit terms of $\a$ in \eqref{eq:angularfull}, and invoking $\o\simeq \m$, the leading-order approximation of the angular part  is obtained  as
\begin{equation}\protect\label{smain}
\left[ \frac{1}{\sin \theta}\frac{\mathrm{d}}{\mathrm{d}\theta}\left( \sin \theta \frac{\mathrm{d}}{\mathrm{d}\theta} \right) -\frac{m^2}{\sin ^2\theta}+\tilde{\lambda}_0\left( \tilde{\lambda}_0-1 \right) \right] S_0=0,
\end{equation}
with $\tilde\lambda_0 \equiv \mu\lambda_{0}$ the dimensionless angular eigenvalue, and $\lambda_0$ and $S_0$ the leading-order value and form of $\lambda$ and $S$.
The solutions are the associated Legendre polynomials $P_{jm}$, if the separation constant satisfies $\tilde\lambda_0 (\tilde\lambda_0 - 1) = j(j + 1)$, so it has two solutions $\tilde\lambda_{0}^+=-j$ and $\tilde\lambda_{0}^-=j+1$.

By substituting the explicit expressions of $B^{\m\n}$ in the leading form of the separable ansatz, and taking the large distance limit $r\gg \l_0$, the spatial components of the vector field simplify to \cite{Baumann:2019eav}
\be
A^i_0 \propto r^{-1} R_0^{\text{far}}(r) \(r\p^iY_{jm}-\l_0Y_{jm}\hat r^i\)\me^{-\mi\o t},
\ee
where $Y_{jm}$ are the scalar spherical harmonics, they relate to the electric vectorial spherical harmonics $Y^i_{j\pm 1,jm}$ via
\be
\1\{\begin{split}
Y^i_{j- 1,jm}&=\frac1{\sqrt{j(2j+1)}}\[r\p^i Y_{jm}+j Y_{jm} \hat r^i\]\\
Y^i_{j+ 1,jm}&=\frac1{\sqrt{(j+1)(2j+1)}}\[r\p^i Y_{jm}-(j+1) Y_{jm} \hat r^i\],
\end{split}\2.
\ee 
by observing and matching the coefficients in front of $Y_{jm}$, we identify $\tilde\l_0^\pm$ correspond to the $j=l\pm 1$ electric modes respectively. Up to this point, the explicit expression of $R_0^{\text{far}}(r)$ is not shown, it should be given by matching different pieces of asymptotic expansions and fixing the boundary conditions, which we will do in the next subsections.

\subsection{Matching three pieces of asymptotic expansions in the radial function}
\protect\label{sec:LOsolutions}

In this subsection, 
we take into account the complete radial range that extends from the BH outer horizon $r_+$ to spatial infinity in the electric modes, by the method of matched asymptotic expansion.
But as mentioned earlier, the additional poles gave rise to the necessity to introduce an intermediate region to match near and far pieces, to obtain the full range behavior.

\deleted{The solutions for the far region, as discussed earlier, exhibit hydrogenic characteristics. }By defining a rescaled radial coordinate as $x\equiv2\sqrt{\mu^2-\omega^2}(r-r_+)$, which mapped the nearer regions to $x\sim \a^2$ and the Bohr radius $r_c\sim \a^{-1}$ to $x\sim 1$, the differential equation in this coordinate reads,
\be
\[\frac{\dif^2}{\dif x^2}+\frac{\n_0}{x}-\frac{\l_0(\l_0+1)}{x^2}-\frac14\]R_0^{\text{far}}=0
\ee
with $\n_0$ the leading order of $\nu\equiv M\mu^2/\sqrt{\mu^2-\omega^2}$, defined in \eqref{nnn};
the solutions are given as
\begin{equation}\protect\label{rfar}
R_{\mathrm{far}}(x)=\mathrm{e}^{-x/2}x^{l +1}U\left(l+1-\nu_0,2+2l ,x \right),
\end{equation}
with $U(...)$ the confluent hypergeometric function of the second kind; for integer values of $\n_0\geq l+1$, it becomes the Laguerre polynomials appearing in the non-relativistic solution \eqref{hr}, which once more suggests the hydrogenic features. Here we omitted possible coefficient in the solution, which is ready to be eliminated by matching to other pieces in the procedure.

For the region near the outer horizon, we define the rescaled coordinate $z\equiv ({r-r_+})/({r_+-r_-})$, representing distance from the outer horizon in unit of $(r_+-r_-)$, and mapping the inner, outer horizon and spatial infinity to $z=-1,0,\infty$. In Eq.~\eqref{eq:originalR}, we can ignore the terms that contain $(r-\hat{r}_{\pm})$ in the denominator and identify $A_{ \pm}$ as $2P_{+}^2$ 
in the small $\a$ approximation. Hence the leading-order radial equation of the near region is
\begin{equation}
\begin{aligned}
&\left[ \frac{\mathrm{d}^2}{\mathrm{d}z^2}+\left( \frac{1}{z}+\frac{1}{z+1} \right) \frac{\mathrm{d}}{\mathrm{d}z}-\frac{j\left( j+1 \right)}{z(z+1)}\right.\left. +\frac{P_{+}^{2}}{z^2}+\frac{P_{+}^{2}}{(z+1)^2}-\frac{2P_{+}^{2}}{z}+\frac{2P_{+}^{2}}{z+1} \right] R^{\mathrm{near}}_0=0,  
\end{aligned}
\protect\label{eq:neareq}
\end{equation}  
the corresponding solutions are known as Kummer’s solutions, with three pairs of equivalent basis denoted by $\{w_1,w_2\}$,$\{w_3,w_4\}$,$\{w_5,w_6\}$ near the poles at $z=-1,0,\infty$ respectively. 
{With physically acceptable purely outgoing solutions at the inner horizon, $w_1$ is used in \cite{Baumann:2019eav}, while the choice of $\{w_5,w_6\}$ is adopted in \cite{Guo:2021xao} to match solutions with far region in scalar case; however, in our vector case here, equation \eqref{eq:neareq} only holds in the near region hence is no longer valid as $z\rightarrow \infty$, and further to better match the boundary condition with reflection near the outer horizon, we identify $\{w_3,w_4\}$ as the proper basis, which is our optimization to the matching procedure, saving from switching between different pairs of basis employed in \cite{Guo:2021xao};} in this regard, the solution reads
\begin{equation}\protect\label{rnear}
    R^{\mathrm{near}}_0(z)=\left( \frac{z}{z+1} \right) ^{\mathrm{i}P_+}\left[ C^{\mathrm{near}}_0w_3\left( z \right) +D^{\mathrm{near}}_0w_4\left( z \right) \right],
\end{equation}
with
\be\1\{\begin{split}
w_{3}(z)&\equiv{}_{2}F_{1}\left(-j,j+1,1+2\mathrm{i}P_{+},-z\right), \\
w_{4}(z)&\equiv(-z)^{-2\mathrm{i}P_{+}}{ }_{2}F_{1}\left(-j-2\mathrm{i}P_{+}, j+1-2 \mathrm{i} P_{+}, 1-2 \mathrm{i} P_{+}, -z\right),
\end{split}\2.\ee
where we have changed the argument $z\rightarrow -z$ to express the basis in terms of hypergeometric function ${ }_{2} F_{1}(...)$.

In order to obtain the full-range radial solution, one must try to extend the near and far region towards each other, which should be achieved by
taking the limits 
$x\rightarrow0$ and $z\rightarrow+\infty$ respectively; 
However, as we've mentioned, the existence of the poles $r=\hat{r}_{ \pm}$ of the radial equation \eqref{eq:originalR} fails to join the two regions, leading to the necessity 
in introduce an intermediate region, which is expressed by defining a rescaled radial coordinate $y\equiv \mu\left(r-r_+\right)$; the equation for the intermediate region is then \cite{Baumann:2019eav}
\begin{equation}\protect\label{rint}
\left[ \frac{\mathrm{d}^2}{\mathrm{d}y^2}+\left( \frac{2}{y}-\frac{2y}{{\tilde{\lambda}_0}^2+y^2} \right) \frac{\mathrm{d}}{\mathrm{d}y}-\frac{\tilde{\lambda}_0\left( \tilde{\lambda}_0-1 \right)}{y^2}-\frac{2\tilde{\lambda}_0}{{\tilde{\lambda}_0}^2+y^2} \right] R_{0}^{\mathrm{int}}=0,
\end{equation}
which has the corresponding solution
\begin{equation}
R_{0}^{\mathrm{int}}(y)=C_{0}^{\mathrm{int}}y^{-\tilde{\lambda}_0}+D_{0}^{\mathrm{int}}y^{-1+\tilde{\lambda}_0}\left[ {\tilde{\lambda}_0}^2\left( 2\tilde{\lambda}_0+1 \right) +\left( 2\tilde{\lambda}_0-1 \right) y^2 \right] ,
\end{equation}
with power indices in polynomials determined by the angular eigenvalues $\tilde\lambda_0 (\tilde\lambda_0 - 1) = j(j + 1)$, as the same in \eqref{smain}.
Since the calculation for the two eigenvalues are parallel, here we only take $\tilde\lambda_0=\tilde\lambda_{0}^+=-j=-(l+1)$ in the following procedure and directly present results for $\tilde\lambda_0=\tilde\lambda_{0}^-=j+1=l$.

To match the three pieces of asymptotic solutions \eqref{rfar}\eqref{rnear}\eqref{rint} in two overlap regions, new matching coordinates should be defined \cite{Baumann:2019eav},
\begin{equation}\1\{
\begin{aligned}
\xi_1 &\equiv \frac{x}{\alpha^\beta}=\frac{2}{\nu} \frac{y}{\alpha^{\beta-1}},\quad  &0<\beta<1, \\
\xi_2 &\equiv \frac{2\mu\left({r}_{+}-{r}_{-}\right)}{\nu} \frac{z}{\alpha^{\beta-1}}=\frac{2}{\nu} \frac{y}{\alpha^{\beta-1}},\quad  &1<\beta<2 ,
\end{aligned}\2.
\end{equation}
with $\xi_1$ connecting $x$ and $y$, and $\xi_2$ connecting $z$ and $y$. 
In these new coordinates, by keeping the leading order in $\a$-expansions and taking the relevant limits of radial coordinates, the behaviors of the near, intermediate and far solutions are given as
\begin{align}\protect\label{rn0m}
\hspace{2em}
    \begin{split}
       R_{0}^{\mathrm{near}}&\sim \ma{C}_{0}^{\mathrm{near}}\left[ 2^{-j}\left( \frac{\alpha ^{\beta -1}\nu}{\mu \left( r_+-r_- \right)} \right) ^j{\xi _2}^j-2^{1+j}\left( \frac{\alpha ^{\beta -1}\nu}{\mu \left( r_+-r_- \right)} \right) ^{-j-1}{\xi _2}^{-j-1}\mathcal{I} _j \right] \\
&+\ma{D}_{0}^{\mathrm{near}}\left[ 2^{-j}\left( \frac{\alpha ^{\beta -1}\nu}{\mu \left( r_+-r_- \right)} \right) ^j{\xi _2}^j+2^{1+j}\left( \frac{\alpha ^{\beta -1}\nu}{\mu \left( r_+-r_- \right)} \right) ^{-j-1}{\xi _2}^{-j-1}\mathcal{I} _j \right];
\end{split}
\end{align}
\vspace{-0.5em}
\begin{numcases}{}
       R_{0}^{\mathrm{int}}\sim C_{\mathrm{int}}2^{-j}\left( \alpha ^{\beta -1}\nu \right) ^j\xi_2 ^j-D_{\mathrm{int}}2^{j+1}j^2\left( 2j-1 \right) \left( \alpha ^{\beta -1}\nu \right) ^{-j-1}\xi_2 ^{-j-1},\protect\label{ri0m1}\\
           R_{0}^{\mathrm{int}}\sim C_{\mathrm{int}}2^{-j}\left( \alpha ^{\beta -1}\nu \right) ^j\xi_1 ^j-D_{\mathrm{int}}2^{j-1}\left( 2j+1 \right) \left( \alpha ^{\beta -1}\nu \right) ^{1-j}\xi_1 ^{1-j};\protect\label{ri0m2}
\end{numcases}
\vspace{-0.5em}
\begin{align}\protect\label{rf0m}\hspace{2em}
\begin{split}
        R_{0}^{\mathrm{far}}&\sim \frac{\Gamma \left( 2j-1 \right)}{\Gamma \left( -\nu \right)}\left( \alpha ^{\beta}\xi _1 \right) ^{1-j}+\frac{\Gamma \left( 1-2j \right)}{\Gamma \left( 1-2j-\nu \right)}\left( \alpha ^{\beta}\xi _1 \right) ^j\\
        &\sim \left( -1 \right) ^{n+1}\delta \nu \left( 2j-2 \right) !n!\left( \alpha ^{\beta}\xi _1 \right) ^{1-j}+\left( -1 \right) ^n\frac{\left( n+2j-1 \right) !}{\left( 2j-1 \right) !}\left( \alpha ^{\beta}\xi _1 \right) ^j.
        \end{split}
\end{align}
where in  \eqref{rn0m} coefficients are, for convenience, rescaled as $\ma{C}_{0}^{\mathrm{near}}\equiv C_{0}^{\mathrm{near}}\Gamma \left( 1+2\mathrm{i}P_+ \right)$ and $\ma{D}_{0}^{\mathrm{near}}\equiv D_{0}^{\mathrm{near}}\Gamma \left( 1-2\mathrm{i}P_+ \right) \mathrm{e}^{2P_+\pi}$, and we used the notation
\begin{align}
        \mathcal{I} _j&\equiv \frac{\left( j! \right) ^2}{\left( 2j \right) !\left( 2j+1 \right) !}\mathrm{i}P_+\prod_{q=1}^j{\left( q^2+4{P_+}^2 \right)},
\end{align}
{which is purely imaginary.
Approximations of Gamma functions around negative integer arguments in \eqref{rf0m} are taken to be the leading terms in Laurent expansions 
$\G(-n+\ep)\sim  {(-1)^n}/({n!\ep})$
 with $n$ a positive integer and $\ep$ an infinitesimal quantity; we see that $\d \n$ only appears in the coefficient of the first term, but is suppressed in the ratio of two infinities in the second.} 
Matching between near and intermediate region goes straightforward by comparing coefficients of $\xi^j$ and $\xi^{-j-1}$ in \eqref{rn0m} and \eqref{ri0m1}, we obtain
\be\1\{\begin{split}
\protect\label{cint1}
C_{\mathrm{int}}&=\left( \ma{C}_{0}^{\mathrm{near}}+\ma{D}_{0}^{\mathrm{near}} \right) \mu ^{-j}\left( r_+-r_- \right) ^{-j},\\
D_{\mathrm{int}}&=\left( \ma{C}_{0}^{\mathrm{near}}-\ma{D}_{0}^{\mathrm{near}} \right) \mu ^{j+1}\left( r_+-r_- \right) ^{j+1}\frac{\mathcal{I} _j}{j^2\left( 2j-1 \right)},
\end{split}\2.\ee
and the parallel process to match the intermediate and far regions results in
\be\1\{\begin{split}
\protect\label{cint2}
C_{\mathrm{int}}&=\left( 2k \right) ^j\mu ^{-j}\frac{\left( n+2j-1 \right) !}{\left( 2j-1 \right) !},\\
D_{\mathrm{int}}&=-\left( 2k \right) ^{1-j}\mu ^{j-1}\delta \nu \left( 2j-2 \right) !n!.
\end{split}\2.\ee
with $k\equiv\sqrt{\mu^2-\omega^2}$ defined for simplicity. Joining \eqref{cint1} and \eqref{cint2}, coefficients in the intermediate region are dropped, and we get relation between the energy level correction $\d \n$ and coefficients in the near region as 
\begin{equation}
\delta \nu =\frac{\ma{C}_{0}^{\mathrm{near}}-\ma{D}_{0}^{\mathrm{near}}}{\ma{C}_{0}^{\mathrm{near}}+\ma{D}_{0}^{\mathrm{near}}}\left( 2k \right) ^{2j-1}\left( r_+-r_- \right) ^{2j+1}\mu ^2\frac{\left( 2j+1 \right)}{\left[ j\left( 2j-1 \right) ! \right] ^2}\frac{\left( n+2j-1 \right) !}{n!}\mathcal{I} _j,
\end{equation}
which holds only for $j=l+1$ as we stated above. For $j=l-1$, the result is given by the similar process as
\begin{equation}
\delta \nu =\frac{\ma{C}_{0}^{\mathrm{near}}-\ma{D}_{0}^{\mathrm{near}}}{\ma{C}_{0}^{\mathrm{near}}+\ma{D}_{0}^{\mathrm{near}}}\left( 2k \right) ^{2j+3}\left( r_+-r_- \right) ^{2j+1}\mu ^{-2}\frac{\left( j+1 \right) ^2}{\left( 2j+1 \right) \left[ \left( 2j+2 \right) ! \right] ^2}\frac{\left( n+2j+3 \right) !}{n!}\mathcal{I} _j.
\end{equation}
These two equations can be consolidated into a unified form for $j=l\pm1$ that reads
\begin{equation}
\protect\label{eq:deltanu}
\delta \nu =\frac{\ma{C}_{0}^{\mathrm{near}}-\ma{D}_{0}^{\mathrm{near}}}{\ma{C}_{0}^{\mathrm{near}}+\ma{D}_{0}^{\mathrm{near}}}\left( 2k \right) ^{2l+1}\left( r_+-r_- \right) ^{2j+1}\mu ^{2\left( j-l \right)}\frac{\left( 2j \right) !\left( 2j+1 \right) !\left( l! \right) ^2}{\left[ j!\left( j+l \right) !\left( j+l+1 \right) ! \right] ^2}\frac{\left( n+2l+1 \right) !}{n!}\mathcal{I} _j.
\end{equation}
Thus the energy level correction $\d\n$ is solely determined by the coefficients in near region, which is further to be identified by the boundary condition at the outer horizon, that we will introduce in the next section.

\section{
The corrected 
spectrums with near-horizon reflection
and superradiant growth anomaly 
}
\protect\label{sec:nearhorizon}
{
\subsection{Reflective boundary conditions near the horizon and corrected superradiance rate}}
Classically, considering the asymptotic behavior of the near radial function, the purely-ingoing boundary conditions at BH horizons follows from the regularity requirements \cite{Berti:2009kk} as
\begin{equation}
\lim _{r_{*} \rightarrow-\infty} R\left(r_{*}\right) \sim \mathrm{e}^{-\mathrm{i}\left(\omega-m \Omega_\pm\right) r_{*}}.
\end{equation}
with the tortoise coordinate $r_{*}$ defined as
\begin{equation}
r_{*}=r+\frac{2 M r_{+}}{r_{+}-r_{-}} \ln \frac{r-r_{+}}{2 M}-\frac{2 M r_{-}}{r_{+}-r_{-}} \ln \frac{r-r_{-}}{2 M},
\end{equation}
where $\O_\pm$ are chosen at $r\ri r_\pm$ respectively and $\O_+\equiv \O_\text{H}$.
{In terms of the coordinate $z$, one can check that in the near region, the purely ingoing conditions are equivalent to $\lim_{z\rightarrow 0}R(z) \sim z^{\mi P_+}$ and $\lim_{z\rightarrow -1}R(z) \sim (z+1)^{-\mi P_+}$, hence $R^\text{near}(z)\sim (\frac z{z+1})^{\mi P_+}$, consistent with equation~\eqref{rnear} with a vanishing $D^\text{near}w_4(z)$ term; the existence of $w_4(z)$ imposes an outgoing term at the outer horizon via $z^{\mi P_+}\ri z^{-\mi P_+}$, while keeps the ingoing condition at the inner horizon fixed.}

{Coincidently, one of the possible consequences of potential microstructures of the horizon, e.g., area quantization \cite{Cardoso:2019apo,Mark:2017dnq}, would}
phenomenologically introduce a reflection parameter $\mathcal{R}$ at the horizon, to encode the modification to a classical BH by including an outgoing term in the boundary condition as  
\begin{equation}
\protect\label{eq:near_codition_with_reflection}
\lim_{r\ri r_+} 
R\left(r_{*}\right) \sim \mathrm{e}^{-\mathrm{i}\left(\omega-m \Omega_\text{H}\right) r_{*}}+\mathcal{R}(\o) ~\mathrm{e}^{\mathrm{i}\left(\omega-m \Omega_\text{H}\right) r_{*}},
\end{equation}
which effectively, in terms of coordinate $z$, 
utilize the basis $w_4(z)\sim z^{-\mi P_+}$ in the near region solution $R^\text{near}(z)$. 
{Note that we always keep the superradiance term $(\omega-m \Omega_\text{H})$ in front of the coordinates, since from the ansatz \eqref{sep_ansatz}, we have
$\me^{-\mi\o t}\me^{\mi m\phi}R(r_*)=\me^{-\mi (\o-m\O_\text{H})t}R(r_*)$ at the horizon,
so this form guarantees the solutions 
\textit{physically} ingoing/outgoing in addition to observed in coordinates \cite{Bardeen:1972fi}. And here we once again emphasize the significance of the quantity $P_+$, with its positivity/negativity denoting a state superradiant or absorptive, while the signs in front denoting a boundary condition ingoing or outgoing; we summarize and list it in Table~\ref{tabel_pplus} for clarity, within which the colored text in left-bottom grid is what we consider in the context.}
\begin{table}[h]
	\centering
	\begin{tabular}{|c|c|c|}
	\hline
	& $P_+>0$ & $P_+<0$\\
	\hline
	$z^{\mi P_+}$ term only & superradiance $\&$  (purely ingong) B.C. & absorption $\&$ (purely ingong) B.C.\\
	\hline
	$z^{-\mi P_+}$ term added  & {\color{violet} superradiance $\&$ (ingong $+$ outgoing) B.C.} & absorption $\&$  (ingong $+$ outgoing) B.C.\\
	\hline
	\end{tabular}
	\caption{States $\&$ boundary conditions (B.C.) in terms of $P_+$. $P_+>0$ or $P_+<0$ denotes superradiance or absorption state; B.C. containing $z^{\mi P_+}$ term only or additional $z^{-\mi P_+}$ term denotes purely ingoing or outgoing included flow of the field.}
	\protect\label{tabel_pplus}
\end{table}

We further assume $\ma R(\o)$ to be real for elegancy, different from that in \cite{Guo:2021xao} by a phase factor;
the spectrum of $\ma R(\o)$ is determined from quantum gravity \cite{Cardoso:2019apo}.
The boundary condition \eqref{eq:near_codition_with_reflection} is to be compared with near region solution \eqref{rnear}, which is manifestly
a linear combination of an ingoing and an outgoing term, weighted by $C_{0}^{\mathrm{near}}$ and $D_{0}^{\mathrm{near}}$ respectively. The reflection parameter then can be extracted by taking the ratio of the coefficients $\mathcal{R} =D_{0}^{\mathrm{near}}/C_{0}^{\mathrm{near}}=\ma D_{0}^{\mathrm{near}}/\ma C_{0}^{\mathrm{near}}\[\Gamma \left( 1-2\mathrm{i}P_+ \right) \mathrm{e}^{2P_+\pi}/\Gamma \left( 1+2\mathrm{i}P_+ \right)\]$, which is brought into \eqref{eq:deltanu} to obtain the form of \eqref{dnn} as
\begin{equation}
    \delta \nu=\delta \bar{\nu} \frac{1-\mathcal{R} \exp \left(2 \mathrm{i} \sum_{q=1}^{j} \phi_{q}\right)}{1+\mathcal{R} \exp \left(2 \mathrm{i} \sum_{q=1}^{j} \phi_{q}\right)}, \qquad\text{with}\quad \phi_{q} \equiv \arctan \frac{2 P_{+}}{q},
\end{equation}
which shares the same form as that in the scalar field case \cite{Guo:2021xao}, but with the orbital angular momentum $l$ replaced by the total angular momentum $j$.
It is straightforward to identify the reflection correction defined in \eqref{dnn} as
\be
\Phi(\ma R)
=\frac{1-\mathcal{R}~\me^{\mathrm{i}\phi_w}}{1+\mathcal{R}~\me^{ \mathrm{i}\phi_w }}=\frac{1-\ma R^2-2\mi \ma R\sin \phi_w}{1+\ma R^2+2\ma R \cos \phi_w}, \qquad \text{with}\quad 
\phi_w=
2\sum_{q=1}^j{\phi _q}.
\ee
By setting $\ma R=0$, we obtain the energy shifts $\del\bar\n$ without reflection, which can be directly read out from \eqref{eq:deltanu} by dropping relevant coefficients, and is observed to be a purely imaginary number in the presence of $\ma I_j$.
Now that we already have $\del\n$ in hand, according to \eqref{odo}, we can immediately get $\del\o$ as
\begin{equation}
   \1\{ \begin{aligned}
        \mathrm{Re}\left( \delta \omega \right) &
        =~~\mi \del\bar\n\frac{M^2\mu ^3}{\bar{n}^3} \text{Im}[\Phi(\ma R)]=-\mi \del\bar\n\frac{M^2\mu ^3}{\bar{n}^3}\frac{2 \ma R\sin \phi_w}{1+\ma R^2+2\ma R \cos \phi_w}
        \\
        \mathrm{Im}\left( \delta \omega \right) &
        =-\mi \del\bar\n\frac{M^2\mu ^3}{\bar{n}^3} \text{Re}[\Phi(\ma R)]=-\mi \del\bar\n\frac{M^2\mu ^3}{\bar{n}^3}\frac{1-\ma R^2}{1+\ma R^2+2\ma R \cos \phi_w}.
    \end{aligned}\2.
\end{equation}
According to \eqref{EG}, the imaginary part $\text{Im}(\del\o)$ directly contributes to superradiant rate $\G$, while the real part $\text{Re}(\del\o)$ is to be compared with $\o_0$; since $\o_0$ is expanded by orders of $\a=M\m$, we extract powers of $M$ and $\n$ in \eqref{eq:deltanu} as, $k=\sqrt{\mu ^2-\omega ^2}\sim \mu \a $, $r_+-r_-=2\sqrt{M^2-a^2}\sim M$ and $P_+\sim \a$, so together we have $-\mathrm{i}\delta \bar{\nu}{M^2\mu ^3}/{\bar{n}^3}\propto(M\mu)^{2l+2j+3}M^2\mu ^3 =\mu (M\mu )^{2l+2j+5}<\mu (M\mu )^5$ when $j\geq1$. Now that energy levels expanded to $\a^5$ is already known \cite{Baumann:2019eav}, we can safely use the result by omitting the shift caused by reflection, which is 
\begin{equation}\protect\label{elvs}
    \begin{aligned}
        E_{\bar{n} l j m}=\mu\left(1-\frac{\alpha^2}{2 \bar{n}^2}-\frac{\alpha^4}{8 \bar{n}^4}+\frac{f_{\bar{n} l j}}{\bar{n}^3} \alpha^4+\frac{h_{l j}}{\bar{n}^3} \chi m \alpha^5+\cdots\right),
    \end{aligned}
\end{equation}
with
\be\1\{\begin{split} 
f_{\bar{n} l j} & =-\frac{4(6 l j+3 l+3 j+2)}{(l+j)(l+j+1)(l+j+2)}+\frac{2}{\bar{n}}\\
h_{l j} & =\frac{16}{(l+j)(l+j+1)(l+j+2)}.
\end{split}\2.\ee
The superradiance rate with reflection is given as
\be\protect\label{g0r}
 {\Gamma}_{\bar njlm}  = {\bar\Gamma}_{\bar njlm} \text{Re}[\Phi(\ma R)]
\ee
with ${\bar\Gamma}_{njlm}$ the rate without reflection,
\be\protect\label{g1r} \begin{split}
 {\bar\Gamma}_{\bar njlm}
 =-\mi \del\bar\n\frac{M^2\mu ^3}{\bar{n}^3}&=\frac{\left( r_+-r_- \right) ^{2j+1}}{M^{2j+2}}\a ^{2l+2j+5} (-\mi\mathcal{I} _j)\\
&\times\frac{2^{2l+2j-2}\left( \bar{n}+l+1 \right) !}{(\bar{n}-l-1)^{2l+4}\bar{n}!}\frac{\left( 2j \right) !\left( 2j+1 \right) !\left( l! \right) ^2}{\left[ j!\left( j+l \right) !\left( j+l+1 \right) ! \right] ^2}\left[ 1+\frac{2(1+l-j)(1-l+j)}{l+j} \right],
    \end{split}\ee
where the last term in the square brackets is also the best educated guess in Ref.~\cite{Baumann:2019eav} for the magnetic modes ($l=j$), so we write it for completeness of the results.

\begin{figure*}[htbp]
    \centering
    \begin{minipage}{0.49\textwidth}
        \includegraphics[scale=0.8]{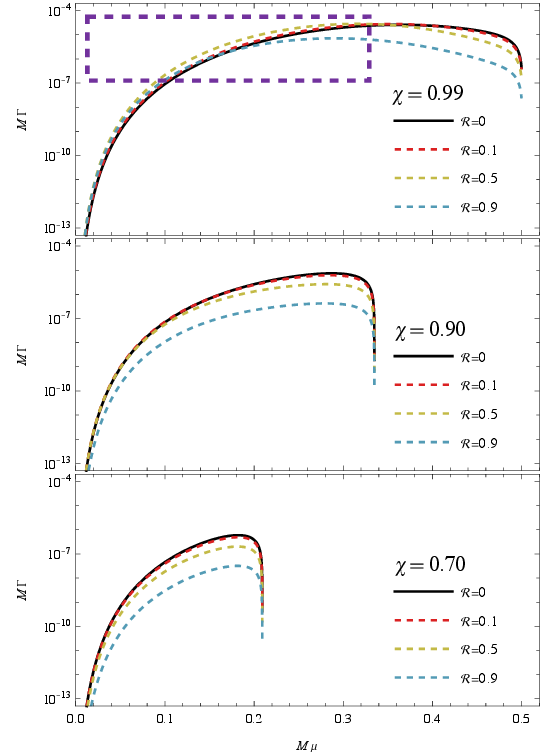}
    \end{minipage}
    \begin{minipage}{0.49\textwidth}
        \includegraphics[scale=0.8]{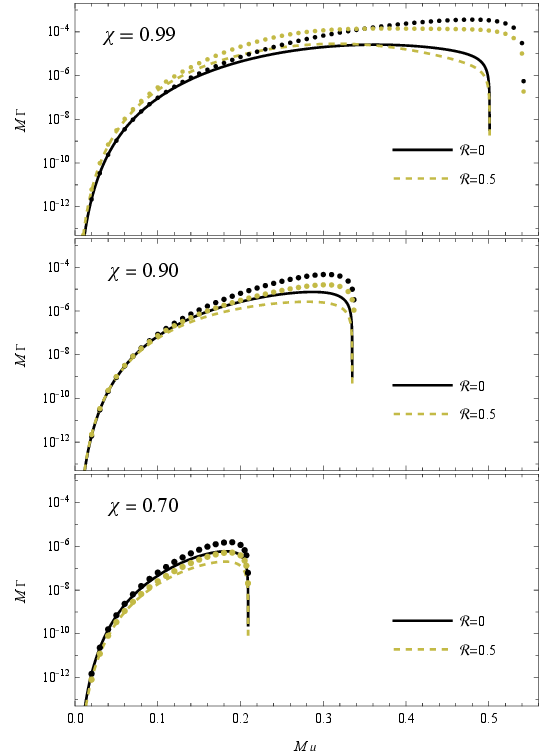}
    \end{minipage}
    \caption{\small Analytic asymptotic (solid and dashed lines) and numerical (dotted lines) results of the dominant superradiant growth rate $\Gamma_{1011}$ as a function of the mass coupling $\a=M\mu$. 
     The BH angular momentum $\chi$ varies: 0.99, 0.9, 0.7 (from top to bottom). The black solid curve denotes cases without reflection. The colored dashed curves (left) and the yellow dotted curves (right) label different reflection parameters $\mathcal{R}$, with the analytic approximation given by Eqs.~\eqref{g0r} and  \eqref{g1r}. The purple dashed box in the left figure marks out the rough region where the anomalous situation occurs: the superradiance rate is initially increased and then decreased when $\ma R$ becomes larger and larger, when the mass coupling is small and BH spin is as high as $\chi=0.99$.
     }
    \protect\label{fig:SR_rates}
\end{figure*}

In this work, we focus on the fastest growing mode in the electric mode, i.e., $\{\bar n,l,j,m\}=\{1,0,1,1\}$, with vanishing cloud angular momentum; since in this mode, the BH spin is totally transferred to the intrinsic spin of the vector field, as we've mentioned in the introduction, making the possible signals most sensitive to the near-horizon geometry of the BH. 
Here we study the effect of the reflection parameter $\mathcal{R}$ on the dominant superradiant growth rate ${\Gamma}_{1011}$ in terms of the mass coupling $\a=M\m$, with different values of BH spin $\chi$, as illustrated 
in Fig.~\ref{fig:SR_rates}. We divide the figure into left and right part: in the left part we plot curves with more values of $\ma R$ to observe the change of the superradiant rate as the reflection parameter varies; in the right part we only plot curves with a zero and a non-zero reflection parameter, but with additional numerical results. The numerical results are obtained using the direct integration method, which is introduced in Ref.~\cite{Dolan:2018dqv}, but here we use the boundary condition \eqref{eq:near_codition_with_reflection} adjusted for reflection. It is seen that the numerical data points lie above the analytical curves, due to the fact that analytical calculations are performed to leading order only, whereas numerical results are computed using the complete expression. It is evident from the comparison that the analytical results are more applicable for cases with small coupling, which is the focus of this study.
\begin{figure*}[htbp]
    \includegraphics[width=\textwidth]{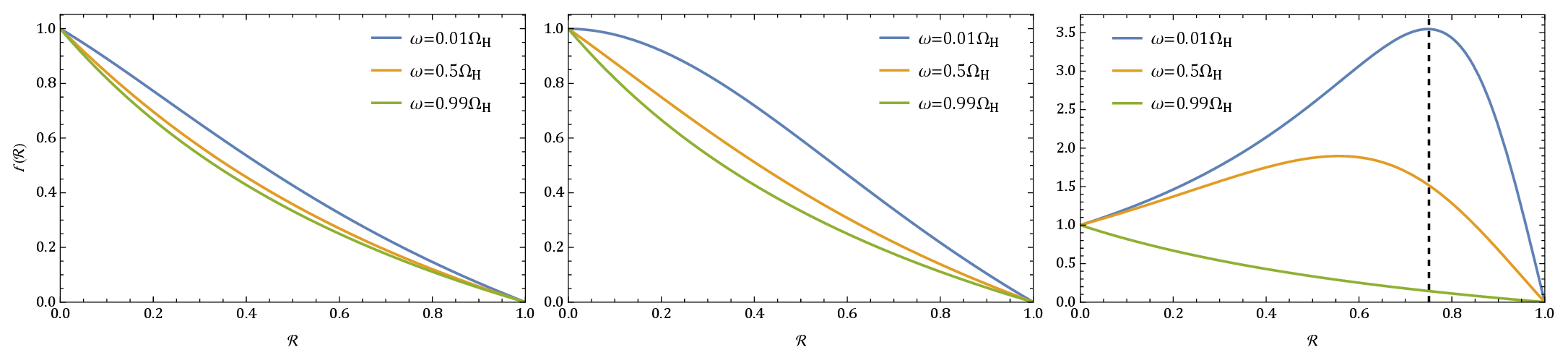}
    \caption{\small The superradiance modifier $f(\mathcal{R})$ as a function of the reflection parameter $\mathcal{R}$  for various $(\chi, \omega)$ values: the BH spin $\chi=0.5$, $\chi=1/\sqrt{2}$, and $\chi=0.99$ cases are shown respectively in left, middle, and right panels, with $\omega=0.01\Omega_\mathrm{H}$ (in blue), $\omega=0.5\Omega_\mathrm{H}$ (in orange), and $\omega=0.99\Omega_\mathrm{H}$ (in green). In the right panel, the critical reflection $\ma R_c\simeq 0.75$
    supporting the highest value of
     $f(\ma R_c)\simeq 3.5$  is labeled by the dashed vertical line.}
    \protect\label{fig:reflection_fR_image}
\end{figure*}

We observe that in the left part, generally the system with a greater reflection parameter has a lower superradiance rate, as shown in the middle and bottom panels; however, an anomalous behavior emerges when the BH spin is high enough (e.g., $\chi=0.99$) and the mass coupling is small, 
as depicted in the top panel: 
an initial increase in the superradiance rate occurs as the reflection parameter rises, then followed by a decrease, which is marked out in the purple dashed box.

\subsection{The superradiance modifier and the growth anomaly conditions}
{
To mathematically investigate this temporary anomalous situation,
we extract the ``superradiance modifier" from Eq. \eqref{g0r} as
\begin{equation}
   f(\mathcal{R}) \equiv  {\Gamma}_{1011} / {\bar\Gamma}_{1011}=\text{Re}[\Phi(\ma R)]_{j=1} = \frac{1-\mathcal{R}^2}{1+\mathcal{R}^2+2\mathcal{R}\cos{(2\phi_1)}},
   \protect\label{eq:fR}
\end{equation}
with $\phi_1=\arctan(2P_+)$.
 Observe that with the increase of $\ma R$, there is a competition between the numerator and the denominator, which is judged by the magnitude of $-\ma R$ and $\cos{(2\phi_1)}$; the anomalous situation occurs when the denominator goes down faster, i.e.,
\be
-\ma R> \cos{[2\arctan(2P_+)]}=\frac2{1+(2P_+)^2}-1,
\ee
and here in case of $m=1$, we use the form of \eqref{pplus} to give
\be\protect\label{ineq}
2P_+=
\qquad
\uwave{\frac{\chi}{\sqrt{1-\chi^2}}\(1-\frac\o{\O_\text{H}}\)>\sqrt{\frac{1+\ma R}{1-\ma R}}}
\qquad
\geqslant 1,
\ee
where the underlined inequality is just the anomalous growth condition. We see that both the BH spin $\chi$ and the superradiance saturation $\o/\O_\text{H}$ play roles here.

Firstly and roughly, given that $1-\o/\O_\text{H}<1$, to make $2P_+>1$, the BH spin $\chi$ should be at least greater than $1/\sqrt2$, otherwise no anomaly can happen, let alone in the presence of increasing $\o/\O_\text{H}$ and non-zero $\ma R$, the BH spin $\chi$ should be even larger, that's why we've only observed the anomaly in the left top panel of Fig.~\ref{fig:SR_rates}; the critical spin $\chi^c$ that anomaly can happen generally depends on the two parameters $\chi^c(\o/\O_\text{H},\ma R)$, with the special case $\chi^c(\o/\O_\text{H}\rightarrow 0,\ma R\rightarrow 0)=1/\sqrt2$ consistent with that we've mentioned. {Later as the system evolves, the superradiance saturation $\o/\O_\text{H}$ goes up and the spin $\chi$ goes down, making the inequality \eqref{ineq} invalid again at some point.
Secondly, given the BH spin, there's a constraint on the mass coupling $\a$, by writing the saturation as $\o/\O\simeq 2\a (1+\sqrt{1-\chi^2})/\chi$, we see that if $\al <\al_\chi\equiv{(\chi-\sqrt{1-\chi^2})}/[{2(1+\sqrt{1-\chi^2})}]$, it would lead to $2P_+<1$, making the anomaly invalid. In high spin case, $\al\in (\a_{\chi=1/\sqrt 2}, \al_{\chi=1})=(0, 0.5)$, there's no anomaly in high mass coupling greater than a half. In particular, for the top panels in Fig.~\ref{fig:SR_rates}, $\al_{\chi=0.99}\simeq 0.37$, it is roughly the turning point for $f(\ma R=0.5)\simeq f(\ma R=0)$, 
To summarize, there are two ingredients necessary to the anomaly: (1) an initial extremely high spin BH; (2) the superradiance is {in the low saturation state.}
\deleted{As an aside, the final value of the modifier when superradiance saturates $\o=\O_\text{H}$ is approxiamated by
$
f(\ma R)|_{\o=\O_\text{H}}=({1-\ma R})/({1+\ma R})
$
,
which will be useful in estimation of the time delay of the whole evolution in Sec.~\ref{subsec:general}
.}
Along with that, 
the turning point $\ma R_c$ where the modifier terminates its boost is calculated by setting $f'(\ma R)=0$ and $f''(\ma R)<0$, we get 
\be\protect\label{rc_new}
\ma R_c=\frac{\sin(2\phi_1)-1}{\cos(2\phi_1)}=\frac{\tan\phi_1-1}{\tan\phi_1+1}=\frac{2P_+-1}{2P_++1}=1-\frac2{1+2P_+},
\ee
which is specified when $\chi$ and $\o/\O_\text{H}$ is given. Numerical values of turning points are shown in Fig.~\ref{fig:reflection_fR_image}, e.g., we use black dashed line to highlight the case where $\mathcal{R}_\mathrm{c}\simeq 0.75$ when $\chi=0.99$ and $\o/\O_\text{H}=0.01$.
It is noted that 
both the high spin and the low saturation are necessary to guarantee the existence and positivity of $\ma R_c$.\\

More elegant geometric forms of the anomaly condition \eqref{ineq} can be obtained, by parametrizing the reflection parameter as $\ma R\equiv \cos(2\phi_{\ma R})$ with $\phi_{\ma R}\in\[0, \pi/4\]$, hence $\sqrt{(1+\ma R)/(1-\ma R)}=\cot \phi_{\ma R}$, and recalling $2P_+=\tan(\phi_1)$, we have
\be\protect\label{ineq2}
\tan\phi_1\tan\phi_{\ma R}>1.
\ee
This can also be achieved by viewing the modifier as square of a ratio between two length,
\be
f(\ma R)
=\frac{1+\mathcal{R}^2-2\ma R\cos(2\phi_{\ma R})}{1+\mathcal{R}^2-2\mathcal{R}\cos{(\pi-2\phi_1)}}
\equiv \frac{\ma T^2}{\ma L^2},
\ee
with
$\ma T=\sin(2\phi_{\ma R})=\sqrt{1-\ma R^2}$ 
the transmission rate at the boundary, and $\ma L$ some kind of a side length of a triangle with other two 
side lengths $1$ and $\ma R$, and an angel 
$\pi-2\phi_1$ between them.
The anomaly holds when $\ma T>\ma L$, i.e., $2\phi_{\ma R}>\pi-2\phi_1$, or simply
\be\protect\label{ineq1}
\phi_{\ma R}+\phi_1>\frac\pi 2,
\ee
which is equivalent to equation~\eqref{ineq2}. 
The geometric forms \eqref{ineq2} and \eqref{ineq1} leads us to a pictorial description of the anomalous growth as shown in Fig.~\ref{geo_anomaly}: The pointers $\vec{\ma R}$ and $\vec{\phi_1}$ are 
``the reflection state" and ``the superradiance state" denoted by arrows pointing at half a unit circle above the horizontal axis, with the former deviated from the positive $x$-axis by the angle $2\phi_{\ma R}$, and the later from the negative $x$-axis by the angle $2\phi_1$; the projection of $\vec{\ma R}$ onto the $x$-axis is the reflection paramter $\ma R$ chosen to be ranged from $0$ to $1$; the sides $\ma T$ and $\ma L$ connect endpoints of $\vec{\ma R}$ and $\vec{\phi_1}$ to the point where the value of $\ma R$ locates. Hence, $\vec{\ma R}$ is restricted in the first quadrant and $\vec{\phi_1}$ is allowed to slide counter-clockwise in the whole upper plane. If the initial superradiance state $\vec{\phi_1}$ is located on the left of the reflection state $\vec{\ma R}$, no anomaly happens; on the contrary, if $\vec{\phi_1}$ is located on the right of $\vec{\ma R}$, anomaly happens in the region where blue and red arcs overlap and then vanishes when $\vec{\phi_1}$ evolves and slides passing through $\vec{\ma R}$.
\begin{figure}[thb]
    \centering
    \includegraphics[scale=0.43,trim=0 150 0 0,clip]{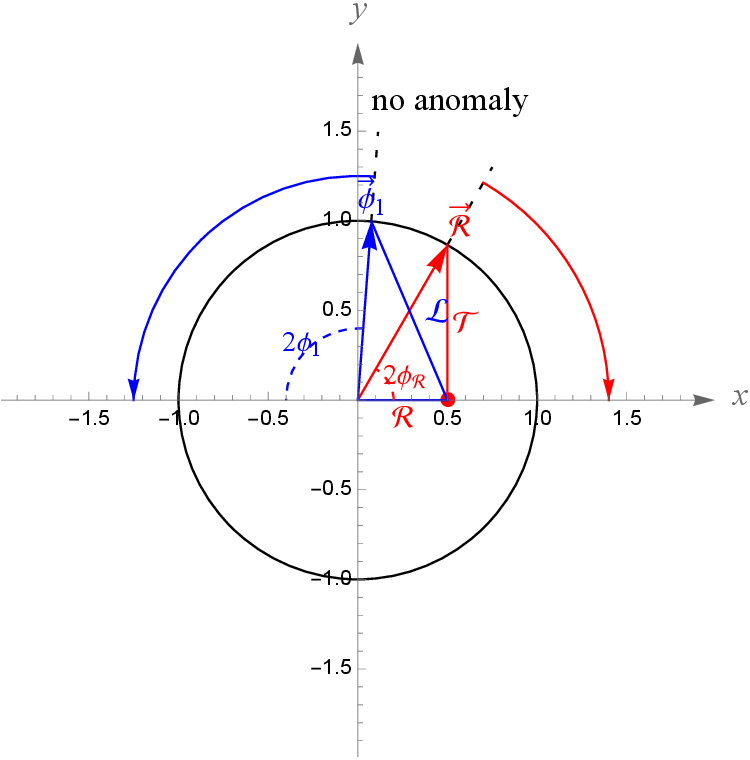}
    \includegraphics[scale=0.43,trim=0 150 0 0,clip]{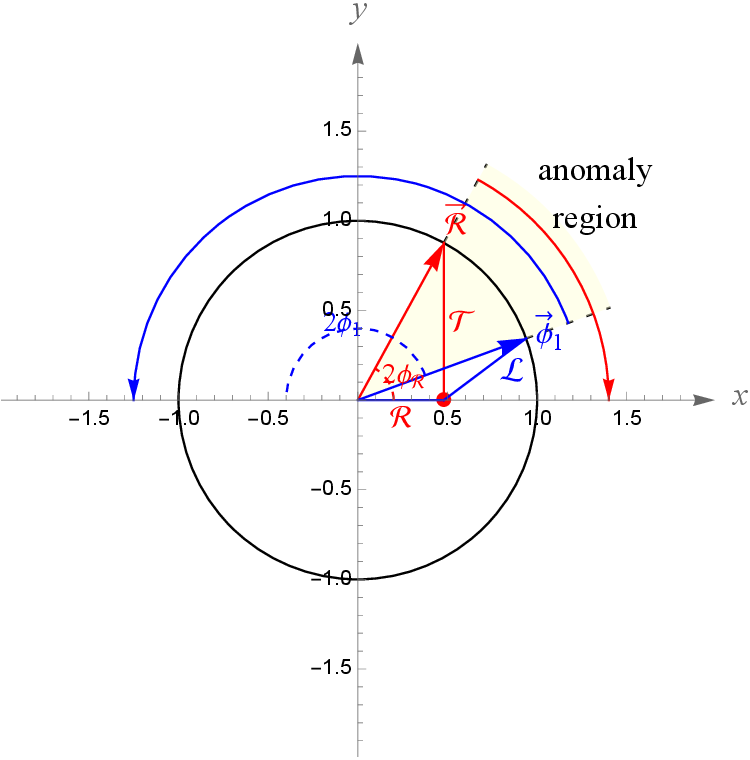}
    \includegraphics[scale=0.43,trim=0 150 0 0,clip]{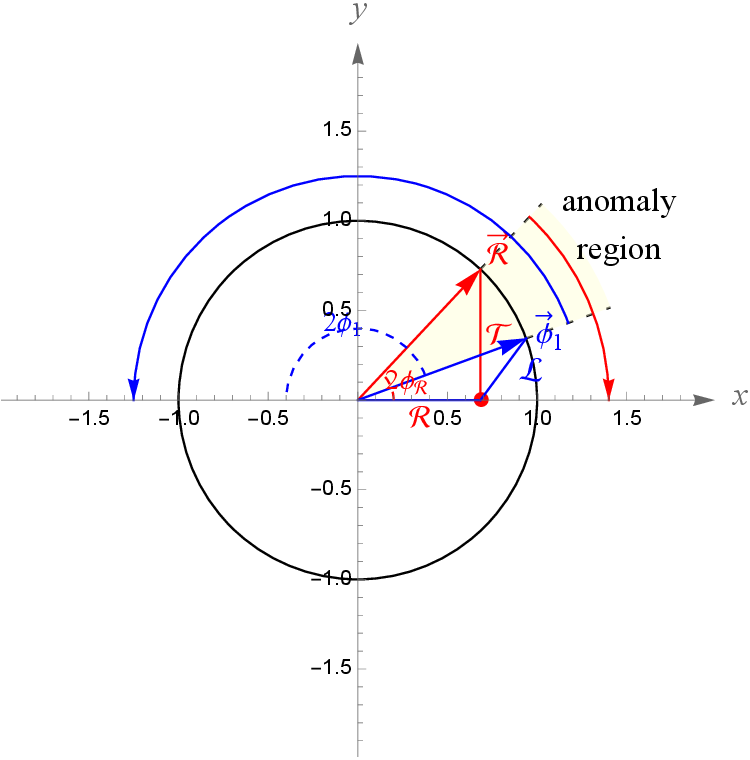}
    \caption{\small Geometrical presentation of cases with different reflection state and superradiance state, denoted by red and blue pointers $\vec{\ma R}$ and $\vec{\phi_1}$ respectively, with the former chosen to be located in the first quadrant while the later sliding dynamically from the first to second quadrant, the corresponding ranges are denoted by red and blue arcs outside the unit circle. The angle $2\phi_{\ma R}$ is between $\vec{\ma R}$ and the positive $x$-axis, while $2\phi_1$ is between $\vec{\phi_1}$ and the negative $x$-axis. The sides $\ma T$ and $\ma R$ are sides facing to $2\phi_{\ma R}$ and $\pi-2\phi_1$ respectively. The left and the middle figures share the same reflection state $\vec{\ma R}$, but with different initial superradiance state $\vec{\phi_1}$; the middle and the right figures share the same different initial superradiance state $\vec{\phi_1}$, but with different reflection state $\vec{\ma R}$.
    The anomaly happens when $2\phi_{\ma R}+2\phi_1>\pi$ (or $\ma T>\ma L$), and vanishes when $2\phi_{\ma R}+2\phi_1<\pi$ (or $\ma T<\ma L$).}
        \protect\label{geo_anomaly}
\end{figure}

Furthermore, by observing the form of \eqref{pplus} and \eqref{difam}, we find
\be
2P_+=\frac{m\O-\o}{\k_+}=-\frac\o{8\pi}\frac{\d A}{\d M}
=\frac1{2\pi k_\text{B}}\frac{\d S_\text{BH}}{\d N},
\ee
with the surface gravity $\k_+=(r_+-r_-)/4Mr_+$ and $\d N=\d M_\text{c}/\o=-\d M/\o$ the growth of  boson occupation number, and further we used the BH entropy $S_\text{BH}=k_B A/4$. In a quasi-adiabatic system containing roughly only the BH and bosonic condensate, $\d S_\text{BH}=-\d S_\text{c}=\d I$, with $I=-S_c$ the information carried away by bosonic condensates, the anomaly condition can be expressed as
\be
\protect\label{anom_cond3}
\frac{\d I}{\d N}>2\pi k_\text{B} \cot\phi_{\ma R},
\ee
which states that growth anomaly happens when information per particle carries exceeds a certain value determined by the reflection parameter. This is more profound than the geometric expressions \eqref{ineq1}\eqref{ineq2}, and may reveal some relations between this anomaly and BH information and microscopic structure. Equation~\eqref{anom_cond3} serve as our most concise expression for the anomaly condition, with physics behind $\ma R$ yet to be implemented in the future work.

We should mention that in spite of different expressions, the anomaly conditions are manifestations of ``rotation-relativistic" effect of a cold ($\chi\simeq 1$) or at most warm ($\chi\gtrsim1/\sqrt{2}\simeq0.7$) BH\footnote{Note that ``hot" and ``cold" BHs are defined from the temperature in BH thermodynamics, which are reverse to the concepts of ``hot" and ``cold" particles by moving velocity.}, whose rotational energy is  leading to or at least comparable to the mass energy, which is analogous to the difference of the kinetic energy ${m_0c^2}\[{(1-v^2/c^2)^{-1/2}}-1\]$ and the rest energy $m_0c^2$ in special relativity \cite{Arvanitaki:2009fg}. The energy stored in spin is a very sensitive function of the spin parameter $\chi$, superradiance turns a BH from a relativistic/cold to a non-relativistic/hot one. 
Possible hints and clearer physics by these expressions deserves further explorations and investigations.

}

\section{Effects of reflection on the BH-condensate evolution and anomaly features}\protect\label{sec:evo}
In this section, we explore the effects of the reflection parameter on the evolution of the BH-condensate system and particularly focus on the case where the anomalous enhancement of superradiant growth is obvious. We provide an overview of the general evolution in Subsection~\ref{subsec:general}, and then focus on the characteristic quantities of the growth anomaly and the induced GW strain deformation in Subsection~\ref{subsec:featured}.

\subsection{
General evolutions and final state estimates
}\protect\label{subsec:general}
In general, the evolution of the BH-condensate system consists of {the following} two \replaced{ingredients}{factors}: the exchange of energy and angular momentum between the BH and the \replaced{contents outside the horizon}{ condensate}, 
 the loss of  energy and angular momentum {of the system} due to GW emissions via annihilation or level transition {in the condensates}, see equation~\eqref{2cl}; we will neglect the transition channel in the following analysis since we  focus on the single vector dominant mode $\{\bar n,l,j,m\}=\{1,0,1,1\}$ only.
{We will also ignore effects of matter accretion and GWs falling towards the horizon in this study.}

 {In this regard, the system is closed by the energy and angular momentum conservation:}
 \be\protect\label{eq:Evoluton_ODE}
 \1\{\begin{split}
    \dot{M}+\dot{M}_{\mathrm{v}}& = -\dot{E}_{\mathrm{GW}},\\
    \dot{J}+\dot{J}_{\mathrm{v}} & = - \dot{J}_{\mathrm{GW}}.
\end{split}\2.
\ee
with $M_\text{v}$ and $J_\mathrm{v}$ the total mass and angular momentum of the vectorial condensate.{The ratio between changes of angular momentum and mass per unit time of the condensates and GWs are just the same as that of a particle \cite{Bekenstein:1973mi}, 
\be
\protect\label{eq:dJH_dt}
\frac{\dot{J}_\mathrm{v}}{\dot{M}_\mathrm{v}}=\frac{\dot{J}_{\mathrm{GW}}}{\dot{E}_{\mathrm{GW}}}=\frac{m}{\omega},
\ee
straightforward as it is, condensates and GWs are due to growth and decays of these particles.}

The growth of the condensates obey
\begin{align}
    \dot{M}_\mathrm{v} & = 2\Gamma M_\mathrm{v}-\G_\text{ann} M_\text{v}^2,
    \protect\label{eq:dEH_dt}
\end{align}
where the superradiant rate $\G$ is defined through equation~\eqref{EG} denoting the changing rate of the field amplitude, and the factor of $2$ comes from the fact that the condensate mass should be relevant to the changing rate of the energy density (or occupation number) \cite{Baryakhtar:2017ngi}; $\G_\text{ann}$ 
is the annihilation 
rate within 
the single energy level $\bar n=1$,
which is much smaller than the typical values of superradiance rate \cite{Arvanitaki:2014wva}.

If the initial condition satisfies $\omega<m\Omega_\mathrm{H}$, the condensate will endure an exponential growth in the superadiant stage, until the saturation when the condensate mass reaches its maximum:
\be
\protect\label{exp_sol}
M_\mathrm{v}^\text{max}\approx M_\text{v}^0 \exp\[\int_{t_0}^{t_f} 2\G(t)\dif t\]\equiv M_\text{v}^0 \exp\[2\overline{\G(t)}\cdot t_\text{grow}\],
\ee
with $M_\text{v}^0$ the initial condensate mass as a seed for the growing, $t_{0,f}$ corresponding to $M_{0,f}$, $t_\text{grow}\equiv t_f-t_0$ the total growth time, and $\overline{\G(t)}$ 
{an effective}
averaged superradiance rate over the integrated time interval. Since $\G(t)$ evolves with time following Equation~\eqref{g_bhav} and Figure~\ref{fig:SR_rates}, generally $t_\text{grow}$ is hard to obtain analytically, while
a commonly used approximation is done by replacing $2\overline{\G(t)}$ by the fastest energy density growth rate as a constant, e.g., $\G_\text{sr}\sim 4\chi\a^6 \m$ for vector clouds \cite{Baryakhtar:2017ngi} leading to the typical superradiance time as $\t_\text{sr}\equiv1/\G_\text{sr}=M/(4\chi\a^7)$, thus a simplified analytical superradiance time is shown as
\begin{align}
   {t_\mathrm{sr}}  \simeq {\t_\text{sr}} \log\1(\frac{M_\mathrm{v}^\text{max}}{M_\text{v}^0}\2), 
\protect\label{eq:tau_sr_0011}
\end{align}
with the typical time $\tau_\text{sr}$ about seconds to minutes for vector clouds around a stellar mass BH, and the exponential growth time $t_\text{sr}$ approximately $2$ orders of magnitude larger due to the logarithm amplification and can last for hours. Straightforward as we just stated, an increase of $2\overline{\G(t)}$ by substitution would certainly underestimate the value of $t_\text{grow}$, so
\be
\t_\text{sr}< t_\text{sr} <t_\text{grow},
\ee
we will directly read $t_\text{grow}$ from numerics afterwards in the next subsection.

During the process throughout, GW emission via annihilations occurs, $\dot E_\text{GW}=\G_\text{ann} M_\text{v}^2$, which is however subdominant in the previous superradiant stage; after the condensate reaches its maximum $\G\ri 0$, it becomes dominant in the evolution Eq~\eqref{eq:dEH_dt}, then the system gradually depletes its energy and undergoes the decay stage. 
The leading term of the GW flux in this stage is 
obtained by fitting numerics to the condensate mass squared proportionality,
which in the small coupling limit reads
\begin{align}
    \lim_{\a \ll 1}\dot{E}_\mathrm{GW}\approx C_\text{GW} \left(\frac{M_\mathrm{v}}{M_f}\right)^2\a_f^{10}.
    \protect\label{eq:GW_single}
\end{align}
where $\a_f\equiv \m M_f$ with $M_f$ the final BH mass at the saturation. The coefficient $C_\text{GW}$ is estimated be be $6.4$ and $60$ in flat and Schwartzchild geometry respectively, we adopt the value of $16.66$ \cite{Siemonsen:2019ebd} in this study, while the latest simulation gives $27.16$ in a recent work \cite{Guo:2024dqd}.
The evolution in this stage is solved by
\be
M_\text{v}=\frac{M_\text{v}^\text{max}}{1+\G_\text{ann}M_\text{v}^\text{max} t},
\ee
with the typical timescale defined to be the ``half-time" of the condensate,
\be
\tau_\text{GW}\equiv \frac1{\G_\text{ann} M^\text{max}_\text{v}}=\frac{M_\text{v}}{\dot E^0_\text{GW}}\approx\frac1{C_\text{GW}}\frac{M^2_\mathrm{f}}{M_\mathrm{v}^\text{max} }\a_f^{-10},
    \protect\label{eq:tau_GW_1011}
\ee
where $\dot E^0_\text{GW}$ is evaluated at $M_\text{v}=M_\text{v}^\text{max}$ in Eq~\eqref{eq:GW_single}.
A direct numerical evaluation shows that $\tau_\mathrm{GW}$ significantly exceeds $t_\text{grow}$, as will be shown in Table~\ref{tab:evo_quantities}. 
Consequently, the depletion of the condensate mass due to GW emission becomes dominant for a much longer timescale than the superradiance stage, consistent with timescales that would be shown in Fig.~\ref{fig:1011_as_0p99}. 

{Back to the superradiant stage, with the suppression of GW emissions, 
analytical approximations can be done} at the 
saturation time,
when
the BH mass and spin of decrease from their initial values $M_0$ and $\chi_0$ to their final values $M_\mathrm{f}$ and $\chi_\mathrm{f}$. {The relations $\D M=\o \D J$ and $\o=\O_\text{H}$ directly yields
\be\protect\label{final1}\1\{\begin{split}
M_\mathrm{v}^\mathrm{max}=M_0-M_\text{f}&=\m \(\chi_0M_0^2-\chi_\text{f}M_\text{f}^2\)\\
\chi_\text{f}&=\frac{4\m M_\text{f}}{1+4(\m M_\text{f})^2},
\end{split}\2.\ee
with $\o\approx \m$ used.
When initial states $(M_0, \chi_0)$ are given, the two unknowns $(M_\text{f}, \chi_\text{f})$ are solved with the above two equations. In the small coupling $\a\equiv M\m< 0.5$ and small energy extraction efficiency $\b\equiv M_\mathrm{v}^\mathrm{max}/M_0<0.3$ limits, the joining equation of the above two,
\be
\b=\a_0(\chi_0-\chi_\text{f})+2\a_0\chi_\text{f}\b-\a_0\chi_\text{f}\b^2
\ee
is simplied by dropping higher order terms to reach
\be\protect\label{final2}
\b=\a_0(\chi_0-4\a_0),
\ee
which solely determines the final state, in a modest accuracy \cite{Zhu:2020tht}. This quantitative approximation holds true for cases with or without reflection,}
while the effect of reflection manifests itself by accelerating or decelerating the superradiant growth. {If the initial state of the bounded system satisfy the anomaly criteria~\eqref{ineq}, }
the reflection parameter enhances the superradiance for a period of time, otherwise the superradiance will be attenuated. The system evolutions are illustrated by solving Eqs.~\eqref{eq:Evoluton_ODE} numerically and results are shown in Fig.~\ref{fig:1011_as_0p99}, and the growth rates of the condensates over time for different BH spins and reflections are shown in Fig.~\ref{fig:1011_as_0p99_velocity}.
\begin{figure*}[htbp]
    \centering
    \includegraphics[width=0.95\linewidth]{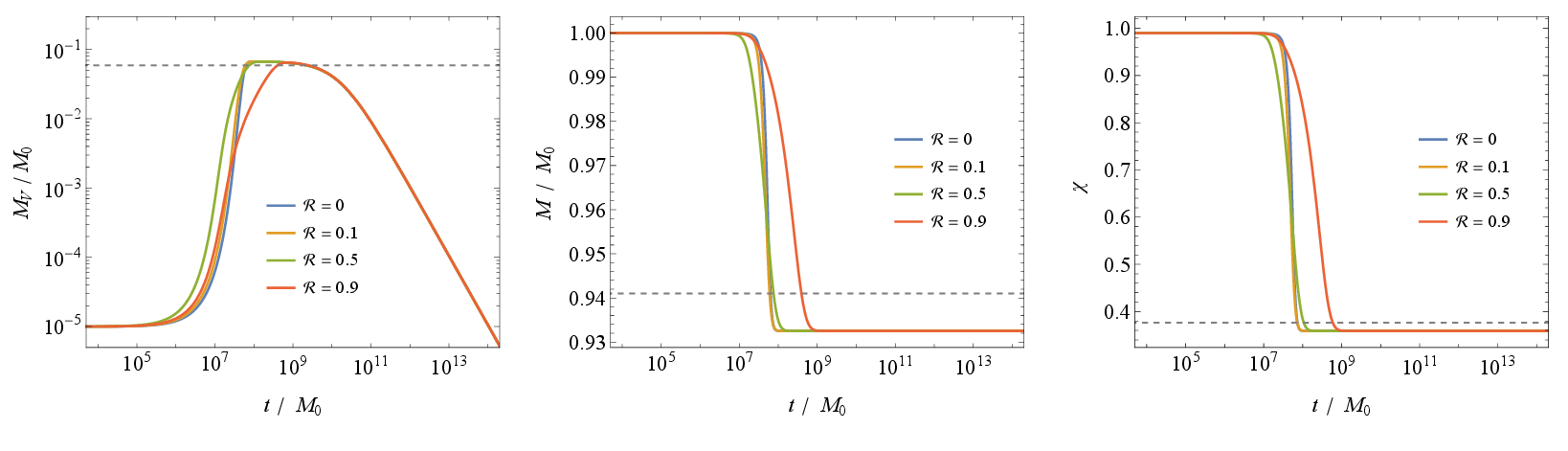}
    \includegraphics[width=0.95\linewidth]{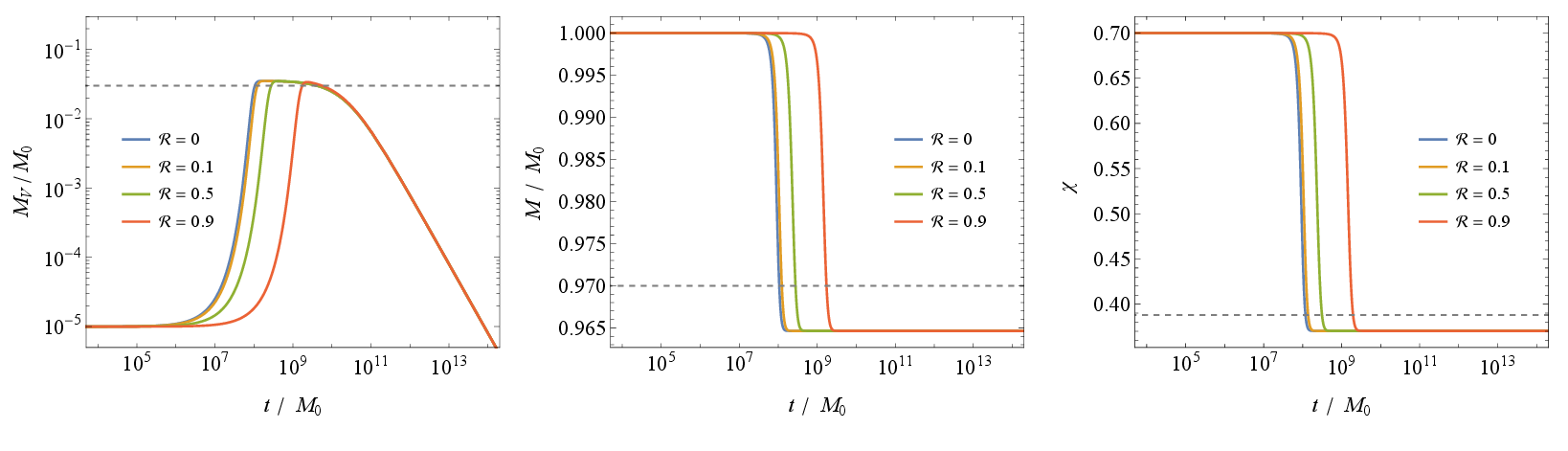}
    \caption{\small The evolution of the BH-condensate system with different reflection parameters $\mathcal{R}$. Only the mode $\{1,0,1,1\}$ is considered in the condensate. The initial BH spins are $\chi_0=0.99$ for the top figure and $0.7$ for the bottom figure. The panels from left to right illustrate the condensate mass, BH mass, and BH spin, respectively. The dashed lines in the panels label the estimations of the maximum condensate masses, the final BH masses, and the final BH spins, i.e. Eqs.~(\ref{final1}-\ref{final2}). The initial mass coupling is $\a_0=0.1$ and {the initial mass of the condensate $M_\text{v}^0$ is set to be $10^{-5}M_0$}. }
    \protect\label{fig:1011_as_0p99}
\end{figure*}
\begin{figure*}[htbp]
    \centering
    \includegraphics[width=0.95\textwidth]{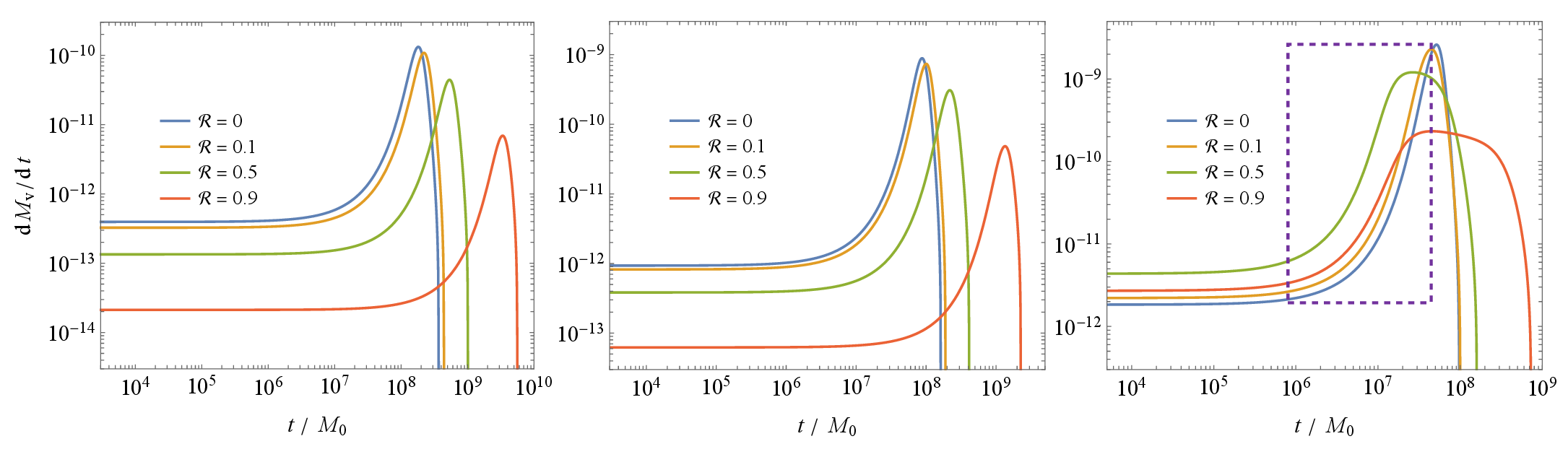}
    \caption{The growth rates of the condensate mass over time for the $\{1,0,1,1\}$ mode. Initial BH spins are $\chi=0.5$ (left panel), $\chi=1/\sqrt{2}$ (middle panel) and $\chi=0.99$ (right panel). The initial mass coupling is $\a_0=0.1$, and the initial condensate mass is $10^{-5}M_0$. The dashed box marks part of the region where growth is faster for all the values chosen of $\mathcal{R}$ when $\chi_0=0.99$.}
    \protect\label{fig:1011_as_0p99_velocity}
\end{figure*}

As we can see from Fig \ref{fig:1011_as_0p99}, for the case in which the BH spin is smaller than the critical spin, e.g., $\chi_0=0.7<1/\sqrt{2}$ (bottom figure),
the superradiant stage is delayed monotonically as the reflection parameter becomes larger; for the extreme case of $\chi_0=0.99$ (top figure), the cases with reflection exhibit accelerated growth rates of the condensate mass at the beginning of the superradiance, and a deceleration in growth rates afterwards.
The comparison of these growth rates is demonstrated more clearly in Fig.~\ref{fig:1011_as_0p99_velocity}, where we denote anomalous enhancement in the purple dashed box.\\



\newpage
However, even if the reflection parameter may enhance the superradiance over some time, the time to reach the \replaced{final state}{critical condition} is always delayed, {as can be seen in Fig.~\ref{fig:1011_as_0p99} the time when $M_\text{v}$ reaches its max, and in Fig.~\ref{fig:1011_as_0p99_velocity} the time when $\dif M_\text{v}/\dif t=0$.} So here we make an interesting metaphor for the anomalous phenomenon we've found, which is described as a race between
``The Tortoise and the Hare":
\begin{wrapfigure}[14]{r}[-10pt]{0.4\textwidth}
  \vspace{-10pt}
    \centering
    \includegraphics[scale=0.2]{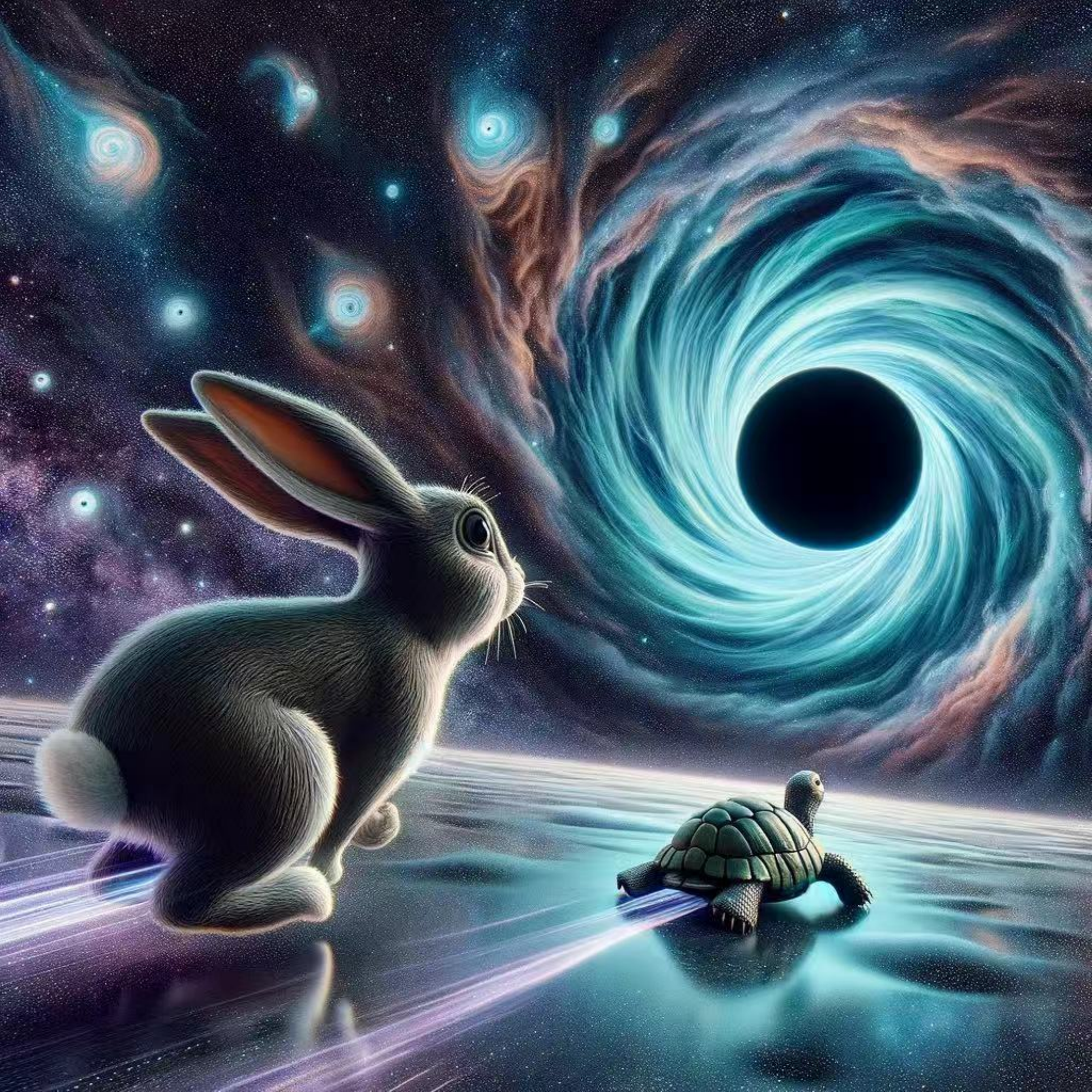}
  \vspace{-22pt}
  \caption{\small A carton animation of the superradiant anomalous growth---``The Tortoise and the Hare", generated by DALL-E, OpenAI.}
  \protect\label{tah}
\end{wrapfigure}
generally as the reflection parameter $\ma R$ becomes larger, the instability rate $\G$ gradually decays to $0$, which intuitively recovers stability as boundary condition tends to be regular, this is the ``tortoise" we originally have; but what we've found in high spin BH case is a temporally running faster ``hare", which brings a larger instability rate as $\ma R$ increases due to Eq.~\eqref{ineq}, but takes a slack afterwards and finally reaches the almost same final state with a time delay (see Fig~\ref{tah} for a carton animation generated by DALL-E, OpenAI). 
In the superradiant stage, this would have impacts on the GW emission power and the GW strain, but with only the later an observable, which is quantified in the next subsection. We will also postulate a physical interpretation for this anomaly in the discussion, which is to be verified/falsified and formulated in the future study.

\subsection{Anomaly features and GW strain magnification}\protect\label{subsec:featured}
In order to gain a more quantitatively description of the reflection-induced growth anomaly, 
\deleted{insightful understanding of the impact of reflection,} in addition to the above quantities that generally describe the evolution, 
we introduce three additional characteristic quantities. These quantities specifically focus on the system with reflection in the anomalous scenario, where reflection amplifies superradiant growth. The first quantity is {$(\varDelta \dot M_\text{v})_{\max}$}, which corresponds to the maximum growth rate difference at a same instant between the reflective and non-reflective system. It demonstrates the utmost extent of reflection to enhance the growth rate throughout the evolution. The second quantity is denoted as $\left( \varDelta M_{\mathrm{v}} \right) _{\max}$, the maximum condensate mass difference between systems with and without reflection. It quantifies the peak advantage in the condensate mass of the reflective system compared to its non-reflective counterpart. The third quantity $t_{\mathrm{adv}}$ gives the period of time during which the reflective system retains its condensate mass advantage, before being overtaken by the non-reflective system.\footnote{Note that by definition, $t_\text{adv}<t_{grow}$ is consistent with numerics in Tab.~\ref{tab:evo_quantities}, but there is possiblity for $t_\text{adv}$ greater than $t_{sr}$ defined in Eq.~\eqref{eq:tau_sr_0011}, which is an analytical approximation.}
In Tab.~\ref{tab:evo_quantities}, we present these quantities
 for the case $\chi_0=0.99$ where the anomaly is most obvious. The \replaced{numerical}{estimated} value of $t_\text{grow}$ indicates that the typical timescales extend with increasing reflection, suggesting that reflection ultimately slows down superradiant growth. On the other hand, both {$(\varDelta \dot M_\text{v})_{\max}$} and $\left( \varDelta M_{\mathrm{v}} \right) _{\max}$ reveal that every reflective system experiences a phase where its growth is enhanced by reflection. These systems maintain a condensate mass advantage for a duration $t_{\mathrm{adv}}$, which decreases as the reflection intensifies.
\begin{table*}[h]
\begin{centering}
\begin{tabular}{|c|c|c|c|c|c|c|c|c|} 
\hline
\multirow{2}{*}{$\chi_0$} & \multirow{2}{*}{$\mathcal{R}$} & \multicolumn{4}{c|}{General evolution quantities}                                                                                                           & \multicolumn{3}{c|}{Anomaly features}                                                                                                                                        \\ 
\cline{3-9}
                          &                                & 
                           $M_{\mathrm{f}}(M_{\mathrm{v}}^{\max }) / M_0$   & $\chi_\text{f}$                & $t_\mathrm{GW} / M_0$               & $t_\text{grow} / M_0$ & {$(\varDelta \dot M_\text{v})_{\max}$} & $\left(\varDelta M_{\mathrm{v}}\right)_{\max}/M_0$ & \multicolumn{1}{l|}{$t_{\mathrm{adv}}/M_0$}  \\ 
\hline
\multirow{4}{*}{$0.99$}   & $0$                            
& \multirow{4}{*}{$0.941(0.059)$} & \multirow{4}{*}{$0.376$} & \multirow{4}{*}{$1.66 \times 10^{10}$} & $9.79 \times 10^7$         & $\text{-}$                                                                          & $\text{-}$                                             & $\text{-}$                                           \\
                          & $0.1$                               
                                                 &                          &                          &                                        & $1.01 \times 10^8$         & $3.23\times 10^{-10}$                                                           & $1.36\times 10^{-2}$                               & $7.70\times 10^{7}$                              \\
                          & $0.5$                                                         &                          &                          &                                        & $1.61 \times 10^8$         & $6.16\times 10^{-10}$                                                           & $2.16\times 10^{-2}$                               & $5.66\times 10^{7}$                              \\
                          & $0.9$                                                        &                          &                          &                                        & $7.36 \times 10^8$         & $4.06\times 10^{-11}$                                                           & $9.36\times 10^{-4}$                               & $3.27\times 10^{7}$                              \\
\hline
\end{tabular}
\par\end{centering}
\caption{\small The general quantities, as well as the characteristic quantities showing the anomaly features for different reflection parameters, with initial mass coupling $\a_0=0.1$, initial condensate mass $M_\mathrm{v,0}=10^{-5}M_0$, and initial BH spin $\chi_0=0.99$. $M_{\mathrm{v}}^{\max}$ 
is the maximum mass of the vector condensate. $M_f$ 
and $\chi_\text{f}$
are the final BH mass and spin. $t_{\mathrm{GW}}$ 
 is the characteristic timescale for the GW emission of the condensate, i.e., the time when the condensate decays to half of $M_{\mathrm{v}}^{\max}$. $t_\text{grow}$ 
 is the duration of superradiant growth of the condensate. {$(\varDelta \dot M_\text{v})_{\max}$} is the maximum difference in growth rate between the systems with and without reflection. $\left(\varDelta M_{\mathrm{v}}\right)_{\max}$ is the maximum difference in the condensate mass between the systems with and without reflection. $t_{\mathrm{adv}}$ is the time span during which the system with reflection maintains its condensate mass advantage before being overtaken by the system without reflection.}
\protect\label{tab:evo_quantities}
\end{table*}

We can also compute the characteristic quantities in SI units, taking a stellar BH with its mass $M_0=100M_\odot$ and the reflection parameter $\mathcal{R}=0.5$ as an example. In this case, the maximum growth rate difference is given by ${(\varDelta \dot M_\text{v})_{\max}}=2.49\times 10^{26}\ \mathrm{kg}\cdot \mathrm{s}^{-1}$, which is approximately $10^{-4}M_\odot$ (about 100 times the mass of the Earth) per second. The maximum difference in condensate mass is $\left(\varDelta M_{\mathrm{v}}\right)_{\max}=4.30\times 10^{30}\ \mathrm{kg}$ or $2.16M_\odot$, and the advantage time is $t_{\mathrm{adv}}=2.79\times 10^{4}\ \mathrm{s}$, or $7.75$ hours {(anomaly signal duration)}. For more massive BHs, the maximum growth rate difference {$(\varDelta \dot M_\text{v})_{\max}$} remains unchanged because the superradiant instability is not solely dependent on $M$. However, both the maximum condensate mass difference $\left(\varDelta M_{\mathrm{v}}\right)_{\max}$ and the advantage time increase $t_{\mathrm{adv}}$ as the BH mass becomes greater.

{
As we've mentioned, the reflection-induced superradiant anomalous growth should have impacts on both GW power and strain, while in a previous numerical work for the scalar case \cite{Guo:2023mel}, the authors pointed out the change in the GW emission rate due to reflection is only obvious for high mass couplings ($\alpha\gtrsim0.5$), which is beyond the superradiant domain ($\alpha\lesssim0.5$); however, they only considered cases as the reflection parameter $\ma R$ close to $0$ or $1$,
for intermediate values in $(0,1)$, where our anomalous growth sets in,
the reflection does have a theoretical possibility to influence the GW power and the GW strain, with the later an observable, thus it's worth formulating and featuring this anomaly.
The GW strain amplitude is expressed in terms of the energy flux as \cite{Arvanitaki:2014wva}
\be
h=\sqrt{\frac{4
}{\o_\text{ann}^2 r^2}\dot E_\text{GW}}
\ee
for a source made up of bosonic condensate annihilating in the corresponding mode, $\o_\text{ann}\simeq 2\m$, emitting power $\dot E_\text{GW}$
 at a distance $r$ away from the Earth.
 In joint with the vector annihilation term, 
 we have 
\be
\protect\label{hc2}
h_\text{v}(t)=M_\text{v}(t)\sqrt{\frac{4
}{\o_\text{ann}^2 r^2} \G_\text{ann}},
\ee
the direct proportionality $h_\text{v}\propto M_\text{v}$ allows us to characterize the anomalous growth of the condensate into the GW strain. Invoking the exponential growth of the condensate, see Eq.~\eqref{exp_sol}, we propose the
  {signal ratio $r_h(\ma R,t)$} 
of the reflection-induced GW strain with that of the case with no reflection,
{\be
r_h(\ma R,t)=\frac{h_\text{v}(\ma R,t)}{h_\text{v}(\ma R=0,t)}=\frac{M_\text{v}(\ma R,t)}{M_\text{v}(\ma R=0,t)}=\exp\1\{\int_{t_0}^{t} 2\bar\G(\ma R,u)\[f(\ma R,u)-1\]\dif u\2\},
\ee}
with $t$ the measured time instant to 
see the {signal ratio}. 
{As such, the featured quantities we proposed can be expressed as
\be\1\{\begin{split}
\frac{\D\dot M_\text{v}(\ma R,t)}{\dot M_\text{v}(\ma R=0,t)}&=f(\ma R,t)r_h(\ma R,t)-1\\
\frac{\D M_\text{v}(\ma R,t)}{M_\text{v}(\ma R=0,t)} &=r_h(\ma R,t)-1,
\end{split}\2.\ee
so we could identify that the time instant for $(\Delta  \dot M_\text{v})_{\max}$ to take place corresponds to the maximum anomolous growth rate $[f(\ma R, t)]_{\max}$ and a still enhancing signal ratio $\dot r_h(\ma R,t)>0$; the time instant for $(\Delta M_\text{v})_{\max}$ to take place corresponds to $f(\ma R, t)=1$ and $[r_h(\ma R,t)]_{\max}$, when the GW strain with reflection is most distinguishable; $t_\text{adv}$ is time when $f(\ma R, t)<1$ and $r_h(\ma R,t)=1$, the GW strains return the same, and anomalous signal ends. So in a time-sequent order, we have 
$
t_{(\Delta  \dot M_\text{v})_{\max }}< t_{\left(\Delta M_{\mathrm{v}}\right)_{\max }}<t_\text{adv}.
$}

{To find the most significant signal ratio with respect to reflection, we assume the measured time 
is short during the evolution, which is consistent with the high BH spin and low saturation conditions to allow anomaly,}
such that
{\be
r_h(\ma R,t)\approx 1+\int_{t_0}^{t} 2\bar\G(u)\[f(\ma R,u)-1\]\dif u,
\ee}
further, as we see in Fig~\ref{fig:1011_as_0p99}, the anomaly is mostly obvious in the fastest growing superradiant stage, 
thus the superradiance rate is set to be 
the fastest rate $2\bar \G(t)\ri \G_\text{sr}$, and time interval is meanwhile selected as $t_\text{H}-t_0\ri \t_\text{sr}$, so 
{\be
r_h(\ma R,t)\approx 1+\G_\text{sr}\t_\text{sr}\[f(\ma R,t)-1\] =f(\ma R,t),
\ee}
where $\G_\text{sr}\t_\text{sr}=1$ is by definition.
We see that the modifier {$f(\ma R,t)$} itself is the magnification of the GW strain, in the earliest stage of the evolution. Meanwhile the GW power $\dot E_\text{GW}$ naturally scales as {$f^2(\ma R,t)$}, which is however not an observable. 
The maximum value of the modifier {$f(\ma R,t)$ with respect to the reflection} takes place at $\ma R_c$, expressed in Equation~\eqref{rc_new}; and we read numerically from Figure~\ref{fig:reflection_fR_image}, when $\chi=0.99$ and $\o/\O_\text{H}=0.01$, the reflection-induced GW strain magnification is about {$r_h(\ma R,t)\approx f(\ma R \simeq 0.75)\simeq 3.5$} times. 

Further, by an analogous analysis, {joining with Eqs.~\eqref{alpha} and \eqref{g_bhav},} the ratio between the strains of vector and scalar/tensor dominant modes with the same reflection reads
\be\label{enhance}
\frac{h_\text{v}(\ma R)}{h_\text{s/t}(\ma R)}\approx\frac{\G_\text{v}}{\G_\text{s/t}}\approx \a^{-2}{\simeq 2500\(\frac{3M_\odot}M\)^2\(\frac{10^{-12}\text{eV}}\m\)^2}
\ee
Just as we stated in the introduction, vector dominant mode serves as the best candidate to characterize effects by reflection, which is mostly obvious in Tab.~I, and Fig.~I. {The reflection-induced anomalous enhancement $r_h$ of the signal strain,} together with the $\a^{-2}$ coupling power magnification for vector case, 
is distinguishable 
if we are lucky enough to catch the moment when a highly spinning Kerr BH captures a small amount of ultralight bosons to begin superradiance, providing an opportunity to shed lights on the micro-structure of the BH horizon. 
}

\section{Conclusion and discussion}
\protect\label{sec:conclusion}
In this study, we investigate the effects of near-horizon reflection of a Kerr BH on the superradiance process and evolution of spin-1 vector bosonic condensate, with the dominant $|1011\>$ mode the most sensitive to horizon geometry and fastest growing mode among the scalar, vector and tensor states, it transfers the BH spin into the intrinsic spin of field rather than the orbital angular momentum, which is a distinguishable feature from the vastly studied scalar case. Hydrogenic structures exhibit in the near-flat spacetime and in the far zone, providing the source for monochromatic GW emissions via energy level transitions and annihilations. In the full analytical treatment, due to the existence of additional poles, three pieces of radial asymptotic solutions should be matched, which is done in the leading order of the angular part; in the solution, we optimize the matching procedure by using a different basis which naturally corresponds to the ingoing and outgoing flux and saves from the switching between different basis as that in many of the previous studies. The near-horizon reflection is denoted by a parameter $\ma R$, which we restrict to be a real value; we studied the modifications due to the reflection parameter on the complex energy level and hence on the superradiance rate, which was plotted in terms of the mass coupling using analytical and numerical results, which have consistent behaviors in the $\a\ll1$ limit. By careful observations, we find that generally as the reflection parameter goes up from $0$ to $1$, the superradiance rate monotonically decreases, while in the early stage of high-spin BH case, the suprradiance rate could temporarily increase with $\ma R$, and then decays to vanish as usual. This peculiar situation also occurs in the scalar cases studied earlier, but lacks a substantial treatment, and it's moreover magnified in the dominant vector case hence deserves a formal analysis; we identified this phenomenon as the ``anomalous growth" of the superradiant condensate, and by singling out the modifier, we figured out the anomaly conditions in three forms: (1) the mathematical form describes well with the figures where we found the situation, which meanwhile gives the critical reflection for the best optimization; (2) the geometrical form goes with two induced angles, which elegantly shows the anomaly in a clock-like diagram with two pointers denoting the superradiance state and the reflection state; (3) the physical form relates the anomaly with the entropy and information carried away by unit superradiant particle, which is a manifestation of rotation-relativistic of fast-spinning BH and deserves deeper investigations in the future once we complement physics of the reflection side. Then, we proceeded with the study of the whole system's evolution, by considering energy and angular momentum exchange among the BH, the condensate and the GWs. We found that although the anomalous growth could temporarily increase the superradiance with reflection, the time for the condensate to reach the maximum is always delayed for a larger reflection parameter. So we make a metaphor for the phenomenon as a race between {``the Tortoise and the Hare", with the former the attenuation of superradiant growth by a higher reflection in a low-spin BH case, and the later the temporarily elevated and then descended growth rate by a higher reflection in a fast-spin BH case.} Further, we proposed three featured quantities to characterize the anomaly: the maximum growth difference, the maximum condensate mass difference, and the advantage time for the anomaly, and we showed the numerics for the case $\chi_0=0.99$ {when} the anomaly is very obvious; in a reasonable numerical estimate, these quantities are considerable in SI units. {We further related the three quantities to the signal ratio of the GW strain, identified the time-sequent order of the corresponding instants; also we checked} the effect of reflection on the GW strain, finding that the modifier is itself the magnification factor { in the early stage of the evolution}, with the best scenario $3.5$ times as the strain without reflection. Although the GW in the earliest superradiance stage is relatively weak, if we were lucky enough to catch a highly-spinning BH capturing a seed of the condensate, it will shed light on the microstructure and quantum nature of the BH horizon.

{
However, along with the potential implications of the three forms of anomaly condition we proposed, a possible physical explanation is yet to be explored. We preliminarily postulate a possibility: the existence of reflection term plays two roles, {one} of which is to occupy the negative energy orbits which leads to attenuating the instability rate, the other is to release negative energy states by modify the BH area thus enlarging the ergosphere volume; the two effects complete with each other, with the former magnificent in relatively slowly rotating BH case, while the later only manifestly shows in the relativistic fast-spinning BH case. Yet, this explanation deserves a further verification/falsification and formulating, which we leave in the future study.


On the other hand, we restricted the reflection parameters to real-valued constants, while introducing detailed {quantum-corrected} models 
could yield more complex effects on superradiant rates and evolution. 
In particular, since superradiant anomaly we've found gives the non-monotonicity dependence on reflection, a specific reflection spectrum would support a certain intermediate value of boson mass, rather than the lightest or heaviest, to produce the most significant GW signal from condensates around high spin BHs. We will report this mass selection procedure along our research line in the future.

The superradiant growth anomaly, is expected to be more magnificent in the tensor polar dipole mode, due to the alleviation of the $\a$-suppression, which is however non-hydrogenic type; and the generalization in the charged case deserves an attentive treatment, which is although less astronomically relevant. The electromagnetic signal features of the growth anomaly is also of potential interest, as it provides wider observational possibilities other than GWs.
Our conclusions may also be extended to more complicated astrophysical environments, such as in cases of accretion, self-interaction, and bosonovas etc. Additional investigations into cases including the presence of a companion star, or considering more types of exotic compact objects (ECOs) as the central body of the system, may also provide comprehensive insights.
 

\section*{Acknowledgements}
\small
We give special thanks to Prof. Hong Zhang and Prof. Shou-Shan Bao for their warm hospitality in Shandong University where part of N. Jia’s work has been done, and give thanks to Mr. Jiangyuan Qian and Dr. Bing Sun for their valuable suggestions and assistance in the early stage of this study. We give additional thanks to Dr. Rong-Zhen Guo and Prof. Qing-Guo Huang for their help in our understanding the asymptotic expansions of some special functions, and give thanks to Prof. Lijing Shao for his expertise in support for our research. The starting work of N. Jia and G-R. Liang is supported by the National Natural Science Foundation of China (Grants~Nos.~12147163 and 12175099), the remaining work of G-R. Liang is supported by Jiangsu Funding Program for Excellent Postdoctoral Talent; the work of Y-D. Guo is supported by the National Natural Science Foundation of China (Grant No. 12075136) and the Natural Science Foundation of Shandong Province (Grant No.~ZR2020MA094); the work of Z-F. Mai is supported by the National Natural Science Foundation of China (Grants~No.~12247128); the work of X. Zhang is supported by the National Natural Science Foundation of China (Grants~Nos.~11975072, 11835009, and 12473001), the National SKA Program of China (Grants~Nos.~2022SKA0110200 and 2022SKA0110203), and the 111 Project (Grant~No.~B16009).

\bibliography{reference}

\end{document}